\documentstyle[psfig]{aa}

\def\ltsima{$\; \buildrel < \over \sim \;$}
\def\simlt{\lower.5ex\hbox{\ltsima}}
\def\gtsima{$\; \buildrel > \over \sim \;$}
\def\simgt{\lower.5ex\hbox{\gtsima}}
\def\cgs{{erg cm$^{-2}$ s$^{-1}$}}
\def\ergs{{erg s$^{-1}$}}
\def\cm2{{cm$^{-2}$}}

\def\xred{{$\chi^{2}_{\rm red}$}}
\def\fhx{{$F_{2-10}$}}
\def\lum{{$L_{2-10}$}}
\def\p1{{Paper I}}
\def\fsx{{$F_{0.5-2}$}}
\def\xmm{{\em XMM--Newton}}
\def\chandra{{\em Chandra}}
\def\nhgal{{N$_{\rm H}^{\rm Gal}$}}
\def\nh{{N$_{\rm H}$}}
\def\chandra{{\em Chandra}}
\def\xmm{{\em XMM--Newton}}
\def\nhgal{{N$_{\rm H}^{\rm Gal}$}}
\def\nh{{N$_{\rm H}$}}  
\def\epic{{\em EPIC}}

\def\pn{{\em PN}}
\def\mos{{\em MOS}}

\def\f14{{10$^{-14}$}}
\def\f13{{10$^{-13}$}}

\begin{document}
\title{An XMM--Newton Study of the Hard X--ray Sky}

\author{E.~Piconcelli\inst{1,2,3}, M.~Cappi\inst{1}, L.~Bassani\inst{1}, G.~Di~Cocco\inst{1}, M.~Dadina\inst{1}}
\institute{IASF/CNR, via Piero Gobetti 101, I--40129 Bologna, Italy
             \and Dipartimento di Astronomia, Universit\`a di
             Bologna, via Ranzani 1, I--40127 Bologna, Italy \and XMM--Newton Science Operation Center/RSSD--ESA, Apartado 50727, E--28080 Madrid, Spain}
\authorrunning{E.~Piconcelli  et al.}
\titlerunning{An {\it XMM--Newton} Study of the Hard X--ray Sky}
\offprints{E.~Piconcelli; e-mail: epiconce@xmm.vilspa.esa.es}
\date{Received / accepted}

\abstract{We report on the spectral properties of a sample of 90 hard X-ray selected serendipitous sources detected
in 12 \xmm~observations with 1 \simlt\fhx\simlt~80 $\times$ 10$^{-14}$ \cgs. 
Approximately 40\% of the sources are optically identified with 0.1 \simlt$z$\simlt~2 and most of them are
classified as broad line AGNs.
A simple model consisting of power law modified by Galactic absorption offers an  acceptable fit to $\sim$ 
65\% of the source spectra. 
This fit yields an average photon index of $\langle\Gamma\rangle \approx$ 1.55 over the whole sample. We also find that the mean slope of the QSOs in our sample turns out to remain nearly constant ($\langle\Gamma\rangle 
\approx$ 1.8--1.9) between 0 \simlt$z$\simlt~2, with no hints of particular trends emerging along $z$.
An additional cold absorption component with 10$^{21}$
\simlt\nh\simlt~10$^{23}$ \cm2 is required in $\sim$ 30\% of the sources.
Considering only subsamples that are complete in flux, we find that
the observed fraction of absorbed sources (i.e. with \nh\simgt~10$^{22}$ \cm2) is $\sim$ 30\%,
with little evolution in the range 2 \simlt\fhx\simlt~80 $\times$ 10$^{-14}$ \cgs. Interestingly, this value is a factor 
$\sim$ 2 lower than predicted by the synthesis models of the CXB.
This finding, detected for the first time in this survey, therefore suggests
that most of the heavily obscured objects which make up the bulk of the CXB will be found at lower fluxes
(\fhx~$<$ 10$^{-14}$ \cgs). 
This mismatch together with other recent observational evidences which contrast with CXB model predictions suggest that one (or more) of the assumptions usually included in these models need to be revised.}

\maketitle 

\keywords{Galaxies: active -- quasars:general --  X--rays: galaxies -- X--rays: diffuse radiation} 


\section{Introduction}
\label{intro}
Deep pencil--beam observations performed with the new generation X--ray telescopes, \chandra~(Mushotzky et al. 2000; Brandt et 
al. 2001; Rosati et al. 2002) and \xmm~(Hasinger et al. 2001), have resolved the bulk of the hard (2--10 keV) cosmic X--ray
background (CXB) into the integrated contribution from discrete sources, pushing the detection limit down to values 
which are several orders of magnitudes fainter than in previous surveys (Moretti et al. 2003). 
Moreover, their excellent angular resolutions have allowed the unambiguos identifications of most 
X--ray sources with active galactic nuclei (AGNs) (Hasinger 2003) providing a unique tool to investigate in detail the formation and evolution of accretion
powered sources (i.e. black holes) over cosmic time as well as the physical connections between nuclear activity and the host 
galaxy (Silk \& Rees 1998; Franceschini et al. 1999).
 
Stacked spectra of these faint sources exhibit a flat slope in full agreement with the spectral shape of the unresolved
CXB in the 2--10 keV band ($\Gamma$ = 1.4; Gendreau et al. 1995), thus resolving the so--called ``spectral paradox'' 
(De Zotti et al. 1982).
Surprisingly, the observed redshift distribution of these sources seems to peak at $z <$ 1 (Gilli 2003, Hasinger 2003), 
at odds with the expectations
of most popular CXB synthesis models (Comastri et al. 1995; Gilli et al. 2001) which predict a peak at $z \sim$ 1.5--2.

Since the diffuse CXB emission is now definitively resolved into point sources, the general interest has 
moved to accurately constrain the physical and evolutionary properties of the different classes of X--ray sources.
However, these extremely deep pencil--beam exposures detect the majority of the sources 
with a very poor counting statistics which prevents from inferring accurate object--by--object X--ray spectral properties: such information 
is nevertheless essential to reveal the physical conditions and the geometry of the matter in the circumnuclear region 
and, hence, how AGNs ultimately work.\\ 

We have therefore undertaken a wide--area search (Piconcelli et al. 2002; hereafter \p1) aimed at exploring the 
individual spectral properties of moderate to faint hard X--ray selected sources. 
The best way to achieve this goal is to  exploit the large collecting area
of \xmm~imaging detectors in order to collect the largest number of targets with X--ray spectra of good quality.
It is worth noting that our survey samples a 2--10 keV flux range (\fhx~$\sim$ 10$^{-14}\div$10$^{-13}$ \cgs) which is mostly 
uncovered by the ultradeep surveys designed to resolve the CXB at much fainter flux levels.
\begin{table*}[!ht]
\caption{Journal of the {\it XMM-Newton} observations.}
\label{tab1}
\begin{center}
\begin{tabular}{lccccccccc}
\hline
\multicolumn{1}{c} {No.} &
\multicolumn{1}{c} {Field} &
\multicolumn{1}{c} {R.A.} &
\multicolumn{1}{c} {Dec.}&
\multicolumn{1}{c} {Date}&
\multicolumn{1}{c} {Obs. ID} &
\multicolumn{3}{c} {Exposure (s)} &
\multicolumn{1}{c} {Filter$^{\dag}$} \\ 
 &                     &           &   &        & &PN   &M1   &M2  &PN M1 M2\\
\hline\hline\\
1&PKS0312$-$770$^{a}$  &03 11 55.0 &$-$76 51 52&2000-03-31 &0122520201&26000&25000&24000&Tc Tc Tc\\
2&MS1229.2$+$6430$^{a}$&12 31 32.0 &$+$64 14 21 &2000-05-20&0124900101&22900&18600&22900 &Th Th Th\\
3&IRAS13349$+$2438$^{a}$&13 37 19.0 &$+$24 23 03&2000-06-20&0096010101&$-$ &41300& 38600 &-- ~Me Th\\
4&Abell 2690$^{a}$&00 00 30.0 &$-$25 07 30&2000-06-01 &0125310101&21000 &16600 &15300 & Me Me Me\\
5&MS 0737.9$+$744$^{a}$&07 44 04.5 &$+$74 33 49& 2000-04-12&0123100101& 15000 &17800 &26100& Th Th Th\\
6&Markarian 205$^{a}$ & 12 21 44.0 &$+$75 18 37&2000-05-07&0124110101&17000 &$-$ &14800& Me ~--~ Me\\
7&Abell 1835$^{a}$&14 01 02.0 &$+$02 52 41 &2000-06-27&0098010101&22900 &23700 &26400 &Th Th Th\\
8&PHL5200$^{b}$       &22 28 30.6  &$-$05 18 32&2001-05-28&0100440101&40200 &34900 &35000 &Tc Tc Tc\\
9&SDSSJ1044$-$0125$^{b}$&10 44 31.8&$-$01 25 09&2000-05-21&0125300101&37500 &28900 &28800 &Th Th Th\\
10 &Lockman Hole$^{b}$&10 52 43.0&$+$57 28 48&2000-04-27&0123700101&33500 &27000 &29900 &Th Tc Th\\
11 &NGC253$^{b}$&00 47 37.1&$-$25 17 41&2000-06-03&0125960101&34100&31600&30300&Me Me Th\\
12 &LBQS2212-1759$^{b}$&22 15 30.9 &$-$17 44 14&2001-11-17&0106660601&80500 &$-$&$-$&Th ~$-$ ~$-$\\
\hline 
\end{tabular} 
\end{center}
$^{\dag}$Optical blocking filters used during observations: Th=thin, Me=medium and Tc=thick.\\
References: $^{a}$ \p1; $^{b}$ this work.
\end{table*}
In \p1 we presented the first results of an initial sample of 41 serendipitous X--ray sources selected from 
seven \xmm~observations with moderate ($\sim$ 20--40 ks) exposures.

In this work, thanks to the addition of further five deeper \xmm~observations (up to $\sim$ 80 ks), 
we extend our detection limit down to \fhx~ $\sim$ 1 $\times$ 10$^{-14}$ \cgs. In this way, we  have been reached values 
closer to the knee of the hard logN--logS distribution,  i.e. where most of the sources accounting 
for the bulk of the CXB reside (Moretti et al. 2003). Accordingly, we are also able to place stronger constraints on some 
input/output parameters of synthesis models of the CXB than in \p1.  
In particular, we are able to provide here a sounder estimate of the fraction of absorbed objects in the faint 
hard X--ray population.
\section{XMM--Newton observations and data reduction}
\label{lab:obs}

The present study is based on a set of 12 observations carried out by the \xmm~satellite (Jansen et al. 2001) from March 2000 to November 
2001, including Performance/Verification phase, Target of Opportunity and Guest Program observations available in the \xmm ~Science Archive.
 The imaging and spectroscopic measurements are taken by the {\it European Photon Imaging Camera} ({\it EPIC}) consisting of one {\it PN} 
back--illuminated CCD array (Struder at al. 2001) and two {\it MOS} front--illuminated CCD arrays (Turner et al. 2001). These fields are 
chosen for their high galactic latitudes ( $\vert b \vert$ $\simgt$ 30 deg) and exposures ($\simgt$ 15 ks) which enable us to collect a large sample of cosmic hard X--ray serendipitous sources without heavy contamination from our Galaxy. 

Table 1 lists the name and the coordinates of the target sources, together with the epoch and the identification number for 
each of the 12  \xmm~observation. Results from the analysis of the first seven observations listed in Table 1 (i.e. from No. 1 to No. 7) 
are reported in \p1.\\ 

For the five newly included \epic~observations (i.e. observations from No. 8 to No. 12 in Table 1) 
we have applied the same data reduction and detection procedures described in \p1.
Similarly, we have included in the final sample only those 49 hard X--ray selected serendipitous sources which satisfy the same selection criterion as used in \p1~(which provides at least \simgt~100 net counts in the 2--10 keV band once all the three {\it EPIC} cameras are taken into account).
The final complete catalog of the 90 X--ray sources is listed in Table 2 together with their \xmm~coordinates.
Approximately 40\% of the sources (i.e. 37 out of 90) are optically identified with 0.1 \simlt$z$\simlt~2 and most of them are classified 
as broad line AGNs.

We further create from this basic sample two subsamples: the BRIGHT sample which includes 42 X-ray sources with\footnote{Fluxes of the sources are calculated using the best fit spectral model (see \p1~and Table 6).} 
\fhx~ $\geq$ 5  $\times$ 10$^{-14}$ \cgs~ and the FAINT sample with 22 sources having 2 $<F_{2-10}<$ 5  
$\times$ 10$^{-14}$ \cgs. This latter sample has been obtained taking into account only the last 5 fields of 
Table~\ref{tab1}, i.e. those with the longest exposures, for which
we could estimate a flux limit of  \fhx~$\sim$ 2 $\times$ 10$^{-14}$ \cgs. This flux limit has been calculated
from the shortest observation, i.e. the Lockman Hole with an exposure of  33.5 ks.
 By doing so, also the FAINT sample is therefore complete down to this flux limit.
\begin{table*}[!t]
\caption{The hard X-ray selected sample.}
\label{tab:sample}
\begin{center}
\begin{tabular}{cccccccc}
\hline
\multicolumn{1}{c} {N} &
\multicolumn{1}{c} {Source name} &
\multicolumn{1}{c} {R.A.}&
\multicolumn{1}{c} {Declination}&
\multicolumn{1}{c} {$z$} &
\multicolumn{1}{c} {Classification}&
\multicolumn{1}{c} {$R$$^{(\dagger)}$} &
\multicolumn{1}{c} {$S_{1.4GHz}$ $^{(\ddagger)}$} \\ 
  & &(J2000) & (J2000)   && &(mag)&(mJy) \\
\hline
\hline
\multicolumn{6}{c}{PKS 0312-770 field}\\
1&CXOUJ031015.9-765131&03 10 15.3& $-$76 51 32 &1.187&BL AGN$^{a}$&17.6 &$-$ \\
2&CXOUJ031209.2-765213&03 12 08.7&$-$76 52 11 &0.89&BL AGN$^{a}$&18.2 &$-$ \\
3&CXOUJ031238.9-765134&03 12 38.8&$-$76 51 31&0.159&Galaxy$^{a}$&17.7 &$-$ \\
4&CXOUJ031253.8-765415&03 12 53.5 &$-$76 54 13 &0.683&Red QSO$^{a}$&22.0 &$-$ \\
5&CXOUJ031312.1-765431&03 13 11.5 &$-$76 54 28 &1.124&BL AGN$^{a}$&18.3 &$-$  \\
6&CXOUJ031314.5-765557&03 13 14.2 &$-$76 55 54 &0.42 &BL AGN$^{a}$&19.1 &$-$ \\
7&XMMUJ030911.9-765824&03 09 11.6 &$-$76 58 24&0.268 &Sey 2$^{b}$&19.1 &$-$ \\
8&XMMUJ031049.6-763901&03 10 49.5 &$-$76 39 01& 0.380& BL AGN$^{b}$&18.6 &$-$ \\
9&XMMUJ031105.1-765156&03 11 05.1 &$-$76 51 56 &$-$ &No cl.&$-$ &$-$ \\
\multicolumn{6}{c}{MS1229.2+6430 field}\\
10&XMMUJ123110.6+641851&12 31 10.6 &$+$64 18 51 &$-$ &No cl.&18.7 &$-$ \\
11&XMMUJ123116.3+641114&12 31 16.3 &$+$64 11 14 &$-$ &No cl.&$-$ &$-$ \\
12&XMMUJ123218.6+640309&12 32 18.6 &$+$64 03 09 &$-$ &No cl.&20.0 &$-$  \\
13&XMMUJ123214.2+640459&12 32 14.2 &$+$64 04 59 &$-$ &No cl.&$-$ &$-$  \\
14&XMMUJ123013.4+642505&12 30 13.4 &$+$64 25 05 &$-$ &No cl.&15.9 &$-$  \\
15&XMMUJ123049.9+640845&12 30 49.9 &$+$64 08 45 &$-$ &No cl.&18.6 &$-$  \\
16&XMMUJ123058.5+641726&12 30 58.5 &$+$64 17 26 &$-$ &No cl.&20.0 &$-$  \\
\multicolumn{6}{c}{IRAS13349+2438 field}\\
17&XMMUJ133730.8+242305 &13 37 30.8 &$+$24 23 05 &$-$ &No cl.&19.39 &$-$  \\
18&XMMUJ133649.3+242004 &13 36 49.3 &$+$24 20 04 &$-$ &No cl.&19.97 &$-$  \\
19&XMMUJ133807.4+242411 &13 38 07.4 &$+$24 24 11 &$-$ &No cl.&18.17 &$-$  \\
20&XMMUJ133747.4+242728 &13 37 47.4 &$+$24 27 28 &$-$ &No cl.&19.5 &$-$  \\
21&XMMUJ133712.6+243252 &13 37 12.6 &$+$24 32 52 &$-$ &No cl.&$-$ &$-$  \\
\multicolumn{6}{c}{Abell 2690 field}\\
22&XMMUJ000031.7-255459&00 00 31.7 &$-$24 54 59 &0.283  &BL AGN$^{b}$&17.7 &$-$ \\
23&XMMUJ000122.8-250019&00 01 22.8 &$-$25 00 19 &0.968  &BL AGN$^{b}$&18.7 &69.2$^{(\star)}$ \\
24&XMMUJ000027.7-250441&00 00 27.7 &$-$25 04 41 &0.335  &BL AGN$^{b}$&18.6 &$-$ \\
25&XMMUJ000100.0-250459&00 01 00.0 &$-$25 04 59 &0.851  &BL AGN$^{b}$&21.9 &130$^{(\star)}$ \\
26&XMMUJ000102.5-245847&00 01 02.5 &$-$24 58 47 &0.433  &BL AGN$^{b}$&20.3 &$-$ \\
27&XMMUJ000106.8-250845&00 01 06.8 &$-$25 08 45 &$-$&$-$&$-$ &$-$ \\
\multicolumn{6}{c}{MS 0737.9+744 field}\\
28&1E0737.0+7436&07 43 12.5 &$+$74 29 35 &0.332 &BL AGN$^{c}$&16.4 &$-$ \\
29&XMMUJ074350.5+743839&07 43 50.5 &$+$74 38 39 &$-$&No cl.&$-$ &$-$ \\
30&1SAX J0741.9+7427&07 42 02.2 &$+$74 26 24 &$-$&No cl.&19.0 &$-$ \\
31&XMMUJ074351.5+744257&07 43 51.5 &$+$74 42 57 &$-$&No cl&20.0&$-$  \\
32&XMMUJ074401.5+743041&07 44 01.5 &$+$74 30 41 &$-$&No cl.&$-$ &87  \\
\multicolumn{6}{c}{Markarian 205 field}\\
33&MS1219.9+7542&12 22 06.6 &$+$75 26 14 &0.238 &NELG$^{d}$&16.91 &$-$ \\
34&MS1218.6+7522&12 20 52.0 &$+$75 05 29 &0.646 &BL AGN$^{d}$ &17.7 &$-$ \\
35&XMMUJ122258.3+751934&12 22 58.3&$+$75 19 34 &0.257 &NELG$^{d}$ &$-$ &$-$ \\
36&XMMUJ122351.3+752224&12 23 51.3&$+$75 22 24 &0.565 &BL AGN$^{d}$ &$-$ &$-$  \\
37&NGC4291&12 20 15.9 &$+$75 22 09 &0.0058&Galaxy$^{d}$ &11.7&$-$ \\
\multicolumn{6}{c}{Abell 1835 field}\\
38&XMMUJ140127.7+025603&14 01 27.7 &$+$02 56 03 &0.265 & BL AGN$^{b}$ &19.7 &1.54$^{(\star)}$\\
39&XMMUJ140053.0+030103&14 00 53.0 &$+$03 01 03 &0.573&BL AGN$^{b}$ &19.7 &$-$ \\
40&XMMUJ140130.7+024529&14 01 30.7 &$+$02 45 29 &$-$ &No cl. &$-$ &$-$ \\
41&XMMUJ140145.0+025330&14 01 45.0 &$+$02 53 30 &$-$$^{\dagger}$ &Galaxy$^{b,\dagger}$ &17.9 &$-$\\
\multicolumn{6}{c}{PHL 5200 field}\\
42&XMMUJ222814.0-051621&22 28 14.0 &$-$05 16 21 &$-$ &No cl. &19.73 &$-$ \\
43&XMMUJ222814.9-052418&22 28 14.9 &$-$05 24 18 &$-$ &No cl. &$-$ &$-$ \\
44&XMMUJ222834.5-052150&22 28 34.5 &$-$05 21 50 &$-$ &No cl.  &$-$ &$-$ \\
45&XMMUJ222822.1-052732&22 28 22.1 &$-$05 27 32 &$-$ &No cl. &$-$ &$-$ \\
46&XMMUJ222850.5-051658&22 28 50.5 &$-$05 16 58 &$-$ &No cl. &$-$ &$-$ \\
47&XMMUJ222823.6-051308&22 28 23.6 &$-$05 13 08 &$-$ &No cl. &$-$ &$-$ \\
48&XMMUJ222905.2-051432&22 29 05.2 &$-$05 14 32 &$-$ &No cl. &$-$ &$-$ \\
49&XMMUJ222732.2-051644&22 27 32.2 &$-$05 16 44 &$-$ &No cl. &19.92 &3.1 \\
\hline
\end{tabular}
\end{center}
\end{table*}
\addtocounter{table}{-1}
\begin{table*}[!t]
\caption{continued}
\begin{center}
\begin{tabular}{cccccccc}
\hline
\multicolumn{1}{c} {N} &
\multicolumn{1}{c} {Source name} &
\multicolumn{1}{c} {R.A.}&
\multicolumn{1}{c} {Declination}&
\multicolumn{1}{c} {$z$} &
\multicolumn{1}{c} {Classification}&
\multicolumn{1}{c} {$R$$^{(\dagger)}$} &
\multicolumn{1}{c} {$S_{1.4GHz}$$^{(\ddagger)}$} \\ 
  & &(J2000) & (J2000)   && &mag&mJy\\
\hline
\hline
50&XMMUJ222826.7-051821&22 28 26.7 &$-$05 18 21 &$-$ &No cl. &$-$ &134\\ 
\multicolumn{6}{c}{SDSS J1044-0125 field}\\
51&XMMUJ104451.3-012229&10 44 51.3 &$-$01 22 29 &$-$&No cl. &$-$ &$-$ \\
52&XMMUJ104445.1-012420&10 44 45.1 &$-$01 24 20 &$-$&No cl. &$-$ &$-$ \\
53&2QZJ104424.8-013520 &10 44 25.1 &$-$01 35 20 &1.57&BL AGN $^{(e)}$&18.61 &$-$\\
54&XMMUJ104509.4-012441&10 45 09.4 &$-$01 24 41 &$-$&No cl. &19.6 &$-$ \\
55&XMMUJ104456.0-012533&10 44 56.0 &$-$01 25 33 &$-$&No cl. &19.0 &$-$ \\
56&XMMUJ104441.9-012655&10 44 41.9 &$-$01 26 55 &$-$&No cl. & $-$&$-$ \\
57&2QZJ104522.0-012845 &10 45 22.3 &$-$01 28 54 &0.782&BL AGN $^{(e)}$&18.5 &$-$ \\
58&XMMUJ104444.6-013315&10 44 44.6 &$-$01 33 15 &$-$&No cl. &18.4 &3.74 \\
\multicolumn{6}{c}{The Lockman Hole field}\\
59&RXJ105421.1+572545  &10 54 21.1 &$+$57 25 45 &0.205&Sey 1.9 $^{(f)}$ &18.3 &0.8 \\
60&RXJ105316.8+573552  &10 53 16.8 &$+$57 35 52 &1.204&BL AGN $^{(f)}$ &19.0 &0.26$^{\star}$ \\
61&RXJ105239.7+572432  &10 52 39.7 &$+$57 24 32 &1.113&BL AGN $^{(f)}$ &18.0 &0.14 \\
62&RXJ105335.1+572542  &10 53 35.1 &$+$57 25 42 &0.784&BL AGN $^{(f)}$ &19.78 &$-$ \\
63&7C 1048+5749        &10 51 48.8 &$+$57 32 48 &0.99 &NL AGN $^{(f)}$ &22.9 &15.39$^{\star}$\\
64&RXJ105339.7+573105  &10 53 39.7 &$+$57 31 05 &0.586 &BL AGN $^{(f)}$ &19.4 &$-$ \\
65&XMMUJ105237.8+573322&10 52 37.8 &$+$57 33 22 &0.707 &NL AGN $^{(f)}$ &22.6 &59.45$^{\star}$\\
66&RXJ105331.8+572454  &10 53 31.8 &$+$57 24 54 &1.956 &BL AGN $^{(f)}$ &19.99 & $-$\\
67&RXJ105350.3+572709  &10 53 50.3 &$+$57 27 09 &1.720 &BL AGN $^{(f)}$ &20.15 &$-$ \\
\multicolumn{6}{c}{NGC 253 field}\\
68&RXJ004759.9-250951  &00 47 59.9 &$-$25 09 51&0.664&BL AGN $^{(g)}$ &17.4 &$-$ \\
69&XMMUJ004722.5-251202&00 47 22.5 &$-$25 12 02&$-$&No cl. &$-$ &$-$ \\
70&RXJ004722.9-251053  &00 47 22.9 &$-$25 10 53&1.25&BL AGN $^{(g)}$ &18.1 &$-$ \\
71&RXJ004647.2-252152  &00 46 47.2 &$-$25 21 52&1.022&BL AGN $^{(g)}$ &20.17 &$-$ \\
72&XMMUJ004818.9-251505&00 48 18.9 &$-$25 15 05&$-$&No cl. &$-$ &$-$ \\
\multicolumn{6}{c}{LBQS 2212-1759 field}\\
73&XMMUJ221536.5-173357&22 15 36.5 &$-$17 33 57&$-$&No cl. &18.0 &$-$ \\
74&XMMUJ221510.7-173644&22 15 10.7 &$-$17 36 44&$-$&No cl. &$-$ &$-$ \\
75&XMMUJ221604.9-175217&22 16 04.9 &$-$17 52 17&$-$&No cl. &20.42 &$-$\\
76&XMMUJ221557.8-174854&22 15 57.8 &$-$17 48 54&$-$&No cl. &$-$ &$-$\\
77&XMMUJ221623.1-174055&22 16 23.1 &$-$17 40 55&$-$&No cl. &$-$ &13.6\\
78&XMMUJ221519.4-175123&22 15 19.4 &$-$17 51 23&$-$&No cl. &$-$ &$-$\\
79&XMMUJ221453.0-174233&22 14 53.0 &$-$17 42 33&$-$&No cl. &$-$ &$-$\\
80&XMMUJ221518.8-174005&22 15 18.8 &$-$17 40 05&$-$&No cl. &$-$ &$-$\\
81&XMMUJ221602.9-174314&22 16 02.9 &$-$17 43 14&$-$&No cl. &$-$ &$-$\\
82&XMMUJ221623.7-174722&22 16 23.7 &$-$17 47 25&$-$&No cl. &20.99 &$-$\\
83&XMMUJ221602.9-174314&22 16 02.9 &$-$17 43 14&$-$&No cl. &$-$ &$-$\\
84&XMMUJ221537.6-173804&22 15 37.6 &$-$17 38 04&$-$&No cl. &$-$ &$-$\\
85&LBQS 2212-1747      &22 15 15.0 &$-$17 32 24&1.159&BL AGN $^{(e)}$&17.3 &$-$\\
86&XMMUJ221623.5-174317&22 16 23.5 &$-$17 43 17&$-$&No cl. &20.9 &$-$\\
87&XMMUJ221550.4-175209&22 15 50.4 &$-$17 52 09&$-$&No cl. &18.1 &$-$\\
88&XMMUJ221533.0-174533&22 15 33.0 &$-$17 45 33&$-$&No cl. &$-$ &$-$\\
89&XMMUJ221456.7-175054&22 14 56.7 &$-$17 50 54&$-$&No cl. &$-$ &$-$\\
90&XMMUJ221523.7-174323&22 15 23.7 &$-$17 43 23&$-$&No cl. &20.44 &$-$\\
\hline
\end{tabular}
\end{center}
Optical classifications and redshifts are taken from: $^{(a)}$ Fiore et al. (2000), $^{(b )}$ Fiore et al. 2003 (F03; in preparation), $^{(c)}$
 Wei et al. (1999), $^{(d)}$ AXIS (e.g. Barcons et al. 2001), $^{(e)}$ Veron--Cetty \& Veron (2001), $^{(f)}$ Mainieri
 et al. (2002), $^{(g)}$ Vogler \& Pietsch (1999). ${(\dagger)}$ There are two possible candidates for the identification of
 this sources: an elliptical galaxy at $z$ = 0.251 or an elliptical galaxy at $z$ =0.254 (F03). $^{(\dagger)}$ Magnitude in the $R$ band.
 Photometric  data are taken from the USNO catalog or F03 whenever available. $^{(\ddagger)}$ Flux density at 1.4 GHz (i.e. 20 cm).
 Data are taken from FIRST and NVSS on--line catalogs. $^{(\star)}$ Radio loud (RL) object. 
\end{table*} 
 \section{Spectral analysis}
\label{Spectral_analysis}
In this Section we focus on the spectral analysis of the X--ray sources selected in the
five new \xmm~observations i.e. sources from No. 42 to No. 90 in Table 2. 
Detailed results about the analysis of the first 41 sources listed in Table 2 can be found in \p1.
We have performed the spectral analysis in the 0.3--10 keV band, choosing the background region in the same detector chip and with the same extraction radius of the source region. 
\begin{table}[!t]
\caption{Spectral fitting results -- I. Fits with a single power law plus Galactic absorption model (SPL).}
\label{tab:SPL}
\begin{center}
\begin{tabular}{cccc}
\hline
\multicolumn{1}{c} {N} &
\multicolumn{1}{c} {$\Gamma$} &
\multicolumn{1}{c} {\xred/(d.o.f.)}&
\multicolumn{1}{c} {Best--fit}\\
 & & & \\
\hline
\hline
\multicolumn{4}{c}{PHL 5200 field (\nhgal = 5.3 $\times$ 10$^{20}$ cm$^{-2}$)} \\
42&1.72$^{+0.16}_{-0.16}$&0.63/(42) &Yes \\
43&1.87$^{+0.44}_{-0.35}$ &1.03/(26) &No \\
44&1.86$^{+0.09}_{-0.09}$ &1.54/(52) &Yes \\
45&1.00$^{+0.20}_{-0.20}$ &0.91/(42)&No \\
46&1.75$^{+0.22}_{-0.22}$ &0.66/(32) &Yes \\
47&1.84$^{+0.12}_{-0.12}$ &0.96/(65) &Yes \\
48&1.00$^{+0.10}_{-0.10}$ &1.45/(74) &No \\
49&1.84$^{+0.09}_{-0.09}$ &1.05/(191) &Yes \\
50&0.85$^{+0.07}_{-0.07}$ &1.35/(176) &No \\
\multicolumn{4}{c}{SDSS J1044-0125 field (\nhgal =4.2 $\times$ 10$^{20}$ cm$^{-2}$)} \\
51&1.28$^{+0.10}_{-0.10}$&1.01/(85) &No \\
52&1.53$^{+0.13}_{-0.13}$&0.99/(44) &Yes \\
53&1.69$^{+0.10}_{-0.10}$&0.93/(86) &Yes \\
54&2.05$^{+0.12}_{-0.12}$&0.84/(88) &Yes \\
55&2.03$^{+0.15}_{-0.15}$&0.85/(37) &Yes \\
56&0.42$^{+0.24}_{-0.24}$&2.10/(19) &No \\
57&2.04$^{+0.12}_{-0.12}$&0.73/(100) &Yes \\
58&0.66$^{+0.28}_{-0.28}$&1.46/(28) &No \\
\multicolumn{4}{c}{Lockman Hole field (\nhgal = 5.5 $\times$ 10$^{19}$ cm$^{-2}$)} \\
59&1.21$^{+0.08}_{-0.08}$&1.54/(361)&No \\
60&1.79$^{+0.05}_{-0.05}$&1.16/(178)&Yes \\
61&2.42$^{+0.10}_{-0.10}$&1.28/(126)&Yes \\
62&2.02$^{+0.08}_{-0.08}$&1.26/(135) &Yes \\
63&0.71$^{+0.19}_{-0.19}$&1.10/(38) &No \\
64&2.34$^{+0.11}_{-0.11}$&1.10/(121) &Yes \\
65&-0.22$^{+0.40}_{-0.49}$&1.07/(11) &No \\
66&1.97$^{+0.14}_{-0.14}$&1.16/(61)&Yes \\
67&1.74$^{+0.28}_{-0.24}$&0.91/(29)&Yes \\
\multicolumn{4}{c}{NGC 253 field (\nhgal = 1.5 $\times$ 10$^{20}$ cm$^{-2}$)} \\
68&1.58$^{+0.09}_{-0.09}$&1.11/(119) &Yes \\
69&-0.24$^{+0.48}_{-0.28}$&0.69/(17) &No \\
70&1.82$^{+0.18}_{-0.18}$&1.00/(69) &Yes \\
71&1.80$^{+0.17}_{-0.17}$&1.29/(85) &Yes \\
72&-0.20$^{+0.48}_{-0.49}$&1.07/(14) &No \\
\multicolumn{4}{c}{LBQS 2212-1759 field (\nhgal = 2.4 $\times$ 10$^{20}$ cm$^{-2}$)} \\
73&2.14$^{+0.27}_{-0.27}$ & 1.10/(63)  &Yes \\
74&-0.18$^{+0.62}_{-0.62}$ & 0.60/(19) &No\\
75&2.08$^{+0.25}_{-0.22}$ & 1.06/(60) &Yes \\
76&1.08$^{+0.44}_{-0.36}$ &0.83/(22) &No \\
77&0.71$^{+0.31}_{-0.31}$ &0.87/(31)&No\\
78&2.10$^{+0.17}_{-0.17}$ &1.15/(62)&Yes \\
79&1.62$^{+0.25}_{-0.24}$ &1.10/(32)&No \\
80&1.10$^{+0.32}_{-0.32}$ &1.02/(29)&No \\
81&2.26$^{+0.11}_{-0.11}$ &1.11/(40)&Yes \\
82&2.03$^{+0.17}_{-0.17}$ &1.08/(133)&Yes \\
83&2.04$^{+0.24}_{-0.24}$ &0.84/(49)&Yes \\
84&0.60$^{+0.28}_{-0.28}$ &1.73/(22)&No \\
85&2.62$^{+0.09}_{-0.10}$ &0.91/(162)&Yes \\
86&1.86$^{+0.15}_{-0.15}$ &0.97/(135)&Yes \\
87&2.22$^{+0.13}_{-0.13}$ &0.82/(88)&Yes \\
88&1.18$^{+0.22}_{-0.22}$ &1.21/(24)&No\\
89&1.06$^{+0.14}_{-0.14}$ &1.29/(63)&No\\
90&2.22$^{+0.11}_{-0.11}$ &0.88/(58)& Yes\\
\hline
\end{tabular}
\end{center}
\end{table}
We adopt as standard a circular extraction region of 35 arcsecs  radius, both for {\it PN} 
and {\it MOS}, shortened if other X--ray sources or CCD gaps are present inside this region.
If during the observation the optical filter of the two {\it MOS} cameras are the same we combine together their spectra. 
Whenever both datasets are available, a joint spectral fitting using the data of both {\it PN} and {\it MOS} is carried out. 
The {\small XSPEC} v.11.0.1 software package has been used to analyse all the background--subtracted source spectra.
In order to permit $\chi^{2}$ fitting, we use a minimum spectral group size of 20 events per data points. 
However, in the case of faint sources with $<$ 400 counts in the broad band 0.3--10 keV, we rebin the data so that there are at least 15
 counts in each bin and we applied the Gehrels weighting function in the calculation of $\chi^{2}$ (Gehrels 1986) since it is a better approximation
in the calculation of $\chi^2$ when the number of net counts is small. 
For the spectral analysis we have used the latest known response matrices and calibration files (January 2002) released by the \xmm~Science 
Operations Centre, taking into account the type of optical filters applied at the top of the telescopes  during the observations
 (see Table~\ref{tab1}). 

Throughout this paper we adopt $H_{0}$ = 50 km s$^{-1}$ Mpc$^{-1}$ and $q_{o}$ = 0 for the calculation of the luminosities. 
Unless stated otherwise, the errors refer to the 90\% confidence level for one interesting parameter (i.e. $\Delta\chi^{2}$ = 2.71; Avni 1976).

\subsection{Spectral fitting}
\subsubsection{Basic models}
\label{basicmodels}

We begin the spectral analysis by fitting the spectra with a simple power law plus Galactic absorption (SPL) model.
This basic spectral parameterization allows us to look for any evidence of absorption and/or  excess emission features: it also provides 
useful indications about the mean slope of the continuum at these hard X--ray flux levels (see Sect.~\ref{cxb}).
Results of these fits are displayed in Table~\ref{tab:SPL}, while the spectrum of each source of the sample together with the relative
data--to--model ratios can be found in Piconcelli (2003). 

We find that the values of \xred~ are statistically acceptable in most cases, thus suggesting that the SPL
model provides a reasonable description of the data for the majority (29 out of 49, i.e. 60\%) of the sources.
 It must be borne in mind
that for the faintest objects, the low quality data prevent a very detailed modeling of some features (i.e. lines, edges) which are
possibly present in their spectra: as consequence the SPL model provides a good description of the overall spectral shape
even if it is not the most appropriate.

Some sources are, however,  clearly not satisfactorily fitted by the SPL model because either the data-to-model ratio residuals are present
 or \xred~$\gg$ 1 (see Table~\ref{tab:SPL}).
Furthermore, spectra with flat photon index ($\Gamma$\simlt~1.3--1.4) indicate the likely presence of intrinsic obscuring
 material which suppresses the soft X--ray continuum.\\

We have therefore refitted each spectrum applying a power law plus an additional absorption component (in source--frame if the redshift is 
known). This spectral model will be indicated as APL hereafter. In Table \ref{tab:APL} we report the relative spectral parameters for 
those source spectra showing a significant improvement at $>$ 95\% confidence level according to an $F-$test once compared to the SPL model fit.
Values of the $F$--statistic and the corresponding significance level are also listed in this Table.

We also include in Table~\ref{tab:APL} three faint objects (i.e. No. 72, No. 74, No. 88) despite the fact that the APL model for them 
is not significantly better than SPL. Indeed their extremely flat SPL spectra strongly suggest the presence of heavy absorption but owing to 
the relatively poor statistics, it is not possible to accurately constrain this component.

In particular,  the application of the APL model to source No.74 reveals strong
obscuration (\nh~$\simeq$ 2 $\times$ 10$^{22}$ cm$^{-2}$) but the relative spectral index still remains very flat and loosely
 constrained because of a likely simultaneous contribution of several unresolved spectral components. 
We therefore fix the photon index $\Gamma$ = 1.9, i.e. the mean value observed in bright AGNs (Nandra \& Pounds 1994), 
to obtain an estimate of the absorption column density value in this source (\nh~$\sim$ 10$^{22-23}$ cm$^{-2}$, see Table~\ref{tab:APL}).

\begin{table}[b]
\caption{Spectral fitting results -- II. Fits with a single power law plus extra absorption component model (APL).}
\label{tab:APL}
\begin{center}
\begin{tabular}{ccccc}
\hline
\multicolumn{1}{c} {N} &
\multicolumn{1}{c} {$\Gamma$} &
\multicolumn{1}{c} {\nh}&
\multicolumn{1}{c} {$F^{(\dagger)}$/C.l.$^{(\ddagger)}$}&
\multicolumn{1}{c} {Best--fit}\\
 & &(10$^{21}$ cm$^{-2}$)&& \\
\hline
\hline
&&&&\\
45&1.61$^{+0.61}_{-0.41}$&3.5$^{+4.0}_{-1.9}$&4.8$/$96\%&Yes\\
48&1.92$^{+0.28}_{-0.23}$&5.4$^{+2.0}_{-1.3}$&70$/>$99.9\%&Yes\\
50&1.09$^{+0.10}_{-0.10}$&1.8$^{+0.9}_{-0.7}$ &15.2$/$98.5\%&No\\
51&1.94$^{+0.14}_{-0.14}$&2.23$^{+0.50}_{-0.50}$&49.6$/>$99.9\%&Yes\\
56&1.74$^{+0.31}_{-0.45}$&11.0$^{+7.10}_{-4.9}$&19$/>$99.9\%&Yes\\
58&1.18$^{+0.49}_{-0.49}$&3.8$^{+4.2}_{-2.1}$&4.5$/$97.5\%&No\\
59&1.85$^{+0.09}_{-0.09}$&2.1$^{+0.3}_{-0.3}$&207$/>$99.9\%&Yes\\
63&1.37$^{+0.44}_{-0.51}$&23.8$^{+20.7}_{-21.0}$&5.3$/$97\%&No\\
65&1.68$^{+0.72}_{-0.87}$&118.5$^{+88.6}_{-109}$&4.8$/$96\%&Yes\\
69&1.63$^{+1.97}_{-1.40}$&5.12$^{+7.88}_{-3.91}$&15.7$/>$99.9\%&Yes\\
72&1.33$^{+2.16}_{-1.77}$&$<$67.5&2.0$/$82\%&No\\
74&1.9f.&31.6$^{+181.8}_{-21.3}$&$-$/$-$&Yes\\
76&1.99$^{+1.58}_{-0.97}$&3.3$^{+7.5}_{-2.7}$&4.9$/$97\%&Yes\\
79&3.03$^{+1.52}_{-0.53}$&3.1$^{+2.0}_{-1.3}$&12.4$/>$99.9\%&No\\
84&1.89$^{+1.31}_{-0.37}$&10.2$^{+17.8}_{-6.2}$&7.6$/$98.5\%&Yes\\
88&1.50$^{+0.46}_{-0.46}$&1.1$^{+1.4}_{-0.8}$&2.70$/$90\%&Yes\\
89&1.91$^{+0.48}_{-0.28}$&3.2$^{+1.5}_{-1.2}$&19.4$/>$99\%&Yes\\
\hline
\end{tabular}
\end{center}
$^{\dagger}$ $F$--statistic value. $^{\ddagger}$ Confidence level with respect to model SPL
 (see Table~\ref{tab:SPL}) using the $F$--statistic.
\end{table}

Using the APL model we find column densities
spanning from $\sim$ 10$^{21}$ to $\sim$ 2 $\times$ 10$^{23}$ cm$^{-2}$ (see Table~\ref{tab:APL}): 
in particular, broad line objects have low amount of cold absorption (i.e., \nh~$<$ 10$^{22}$ \cm2), similarly to what found in \p1.
Interestingly both 7C~1048$+$5749 and XMMU~J105237.8+573322 (source No. 63 and No. 65, respectively)
show X--ray luminosities exceeding 10$^{44}$ erg/s as well as a column density \nh~$>$ 10$^{22}$ \cm2, thus becoming candidates to be 
type 2 QSOs (e.g. Sect.~\ref{HXSQs}, Mainieri et al. 2002). Note that
the values of spectral parameters $\Gamma$ and \nh~derived by our analysis 
of the  sources in the Lockman Hole field (i.e. sources from No. 59 to No. 67) fully agree with 
those obtained by Mainieri et al. (2002) using a longer ($\sim$ 100 ks) \xmm~observation.

As expected the introduction of an additional absorption component produces a significant
steepening of the continuum slope in most of the sources listed in Table~\ref{tab:APL}. 
However, a sizeable number of objects (5 out of 17) still have flat spectra with $\Gamma$\simlt~1.2 
and/or exhibit residuals in their data--to--model ratios. 
Thus a further and more detailed analysis has been carried out in order to take into account also these additional spectral features 
(see Sect.~\ref{morecomplex}).\\

Finally, as already done in \p1, we also fit all the spectra with the APL model fixing $\Gamma$ = 1.9 (and
$z$ = 1 for the optically unidentified sources) in order to overcome a possible underestimation of the intrinsic 
column densities in those sources with the lowest statistics
and/or without redshift information\footnote{The effective column density
N$_{\rm H}^{eff}$ has  the following redshift dependance: N$_{\rm H}^{eff} \propto$ \nh(1 + $z$)$^{2.6}$ (Barger et al. 2002).}. 
We choose $z$ = 1 on the basis of the findings reported in recent optical follow-ups of ultradeep X--ray surveys 
(Hasinger et al. 2003; Cowie et al. 2003) which suggest a peak  at $z$ \simlt~1 in the redshift distribution
of the sources making the CXB. Results of this spectral fitting will be discussed in Sect.~\ref{cxb} in the frame of the observational
constraints on the predictions of the synthesis model of the CXB.
\subsubsection{More complex models and peculiar sources}
\label{morecomplex}
\begin{figure}[ht]
\begin{center}
\psfig{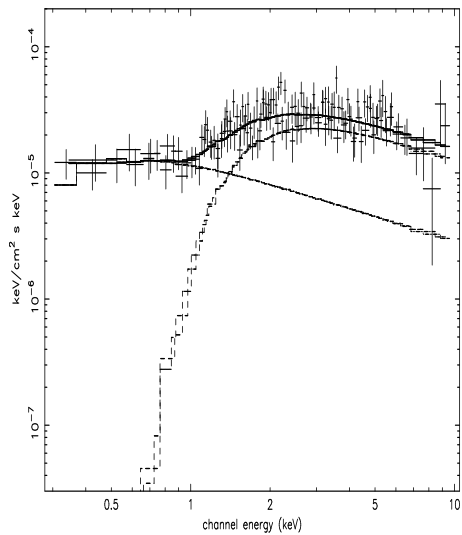}
\caption{The \epic~(\pn~and \mos) spectrum of the unidentified source No.~50 in the PHL 5200 FOV with the soft--excess component
fitted with a scattered power law model (PL in Table~\ref{tab:complex}).}
\label{fig:50}
\end{center}
\end{figure}

Although intrinsic absorption suppresses a sizeable fraction of the soft X--ray primary continuum, 
many Type 2 AGNs are characterized by a soft--excess component which is either originating in a circumnuclear diffuse starburst 
and/or is due to reprocessed emission scattered along our line of sight by a photoionized gas located just above the obscuring torus
(Turner et al. 1997, Kinkhabwala et al. 2002). This is the case for seven X--ray sources in our sample 
(i.e. Nos. 50, 58, 63, 72, 77\footnote{Source No. 77 has been included here due to its very flat photon index derived by the SPL model.
Accordingly, this source is likely obscured by a large amount of absorption but the data quality does
 not allow a more accurate spectral modeling.}, 79\footnote{Although the statistical improvement is not so significant source No. 79 has been included here because after the introduction of a soft-excess component the resulting photon index is $\Gamma \approx$ 2 (i.e. a value commonly found in AGNs) instead of the unusual steep slope derived by the APL model ($\Gamma \approx$ 3)} and 80; see Table~\ref{tab:APL}), for 
which we  have therefore included such a component (significant at \simgt~90\% confidence level) in their best fit model. A thermal Raymond--Smith component (labelled with TM in Table~\ref{tab:complex}) is  required in source Nos. 79 and 80; 
while  an additional power law (labelled with PL in Table~\ref{tab:complex}) is added for the latter five
remaining objects. 
The metallicity of the thermal component is fixed to the solar value while  
the spectral index of the second power law is put equal to the value found for the hard X--ray primary power law,
as expected in the case of a scattered component (Turner et al. 1997).
An example of an absorbed source (No.~50) for which we have applied an additional power law spectral parameterization is shown 
in Figure~\ref{fig:50}.

The resulting average increase of the intrinsic column density value due to the addition of a soft excess component is
$\langle$\nh$\rangle$ $\approx$ 10$^{22}$~\cm2.
\begin{table*}[!t]
\caption{Spectral fitting results -- III. Fits with complex models.}
\label{tab:complex}
\begin{center}
\begin{tabular}{ccccccc}
\hline\hline
\multicolumn{1}{c} {Source No.~} &
\multicolumn{1}{c} {Model$^\dagger$}&
\multicolumn{1}{c} {$\Gamma$} &
\multicolumn{1}{c} {$N_{\rm H}$}&
\multicolumn{1}{c} {E$_{edge}$/$k$T}&
\multicolumn{1}{c} {$F-$statistic} &
\multicolumn{1}{c} {C.l.$^\ddagger$}\\
&&&(10$^{21}$ cm$^{-2}$)&(keV)&&\\
\hline
 &&                      &                    &   &      &\\
43&WA&2.08$^{+0.36}_{-0.31}$&$\equiv$\nhgal&0.55$^{+0.08}_{-0.07}$&4.7&96\%\\
50&PL&1.66$^{+0.05}_{-0.05}$&14.3$^{+3.3}_{-3.3}$ &$-$&48.8&$>$99.9\%\\
58&PL&1.41$^{+0.26}_{-0.57}$&7.4$^{+6.7}_{-5.9}$  &$-$&2.8&90\%\\
63&PL&1.64$^{+0.43}_{-0.25}$&60$^{+59}_{-51}$     &$-$&3.3&93\%\\
72&PL&1.90f.&53.3$^{+50.4}_{-33.9}$&$-$&4.6&95\%\\
77&PL&1.33$^{+0.63}_{-0.87}$&7.7$^{+20.1}_{-5.2}$&$-$&2.1&85\%\\
79&TM&2.06$^{+1.19}_{-0.55}$&2.9$^{+6.9}_{-2.1}$&0.43$^{+3.62}_{-0.23}$&0.4&35\%\\
80&TM&1.90f.& 7.8$^{+6.7}_{-4.3}$&0.15$^{+0.14}_{-0.10}$&2.6&90\%\\
\hline
\end{tabular} 
\end{center}
$^\dagger$ PL$=$SPL (or APL) + non-thermal model for the soft excess component; TM$=$SPL (or APL) + thermal model
for the soft excess component; WA$=$SPL (or APL) + warm absorber features.     
$^\ddagger$ Confidence level with respect to model SPL (see Table~\ref{tab:SPL}) or model APL (see Table~\ref{tab:APL}) using the $F$--statistic.
\end{table*}

Finally, a clear warm absorber signature is present in the SPL spectrum of  the unidentified source No.~43:
we have therefore added in its fitting  model an absorption edge to parameterize this feature.
The improvement in the $\chi^2$ is significant at \simgt 96\% confidence level with a resulting observed--frame energy{\footnote{Assuming that the observed edge is due to OVII(OVIII), we infer a redshift $z$ = 0.35(0.81) for this X--ray source.} for 
the edge E$_{edge}$ = 0.55$^{+0.08}_{-0.07}$ keV, likely due to highly ionized OVII/OVIII as commonly found in many Seyfert 1s (Reynolds 1997). 

In Table~\ref{best_flux2} flux in the 0.5--2 keV and 2--10 keV band, 2--10 keV luminosity and best fit model are listed for
all sources presented in this Section. 
\section{Results on the whole sample}
\label{results}
Adding the five \xmm~exposures presented in Sect.~3 (i.e. observations Nos. 8 to 12 in Table 1), 
the number of hard X--ray selected sources in our sample has increased from 41 (in \p1) to 90.
In this Section we present the results obtained by taking into account the whole sample as listed in Table~2.
The present \xmm~observations yield the first 0.3--10 keV spectrum of a large fraction of the X--ray sources investigated 
in this work, since most of them have not been detected by previous less sensitive X--ray telescopes.
\begin{figure}[b]
\begin{center}
\psfig{file=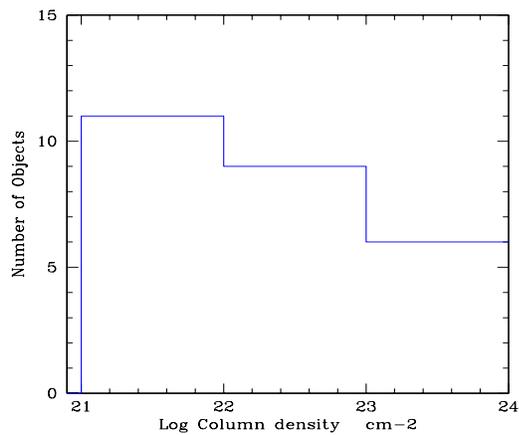,width=7cm,height=6cm}
\caption{Distribution of the absorption column densities for the 26 sources in the sample requiring
the introduction of a significant absorption component (see Table~\ref{tab:APL}, Table~\ref{tab:complex}; and  Table 4 and 5 in \p1).}
\label{fig:absorption}
\end{center}
\end{figure}
\begin{table}[!t]
\label{best_flux2}
\caption{\xmm~properties of the 49 X--ray sources. Fluxes and luminosities are calculated for each source using the best--fit model
listed in column 2 (see Table~\ref{tab:complex} for details).}
\label{best_flux2}\begin{center}
\begin{tabular}{ccccc}
\hline
\multicolumn{1}{c} {N} &
\multicolumn{1}{c} {Best--fit} &
\multicolumn{1}{c} {$F_{0.5-2}^{(a)}$}&
\multicolumn{1}{c} {$F_{2-10}^{(a)}$}&
\multicolumn{1}{c} {$L_{2-10}^{(b)}$}\\
 &Model&& &\\
\hline
\hline
42&SPL &2.08&   3.79&$-$\\              
43&WA  &1.21&   1.69&$-$\\              
44&SPL &2.73&   4.91&$-$\\
45&APL &1.78&   4.38&$-$\\    
46&SPL &1.63&   2.91&$-$\\      
47&SPL &3.18&   5.01&$-$\\
48&APL &3.07&   10.61&$-$\\     
49&SPL &1.73&   29.70&$-$\\     
50&PL  &4.39&   28.1&$-$\\
51&APL &3.83&   9.11&$-$\\      
52&SPL &1.63&   4.33&$-$\\      
53&SPL &4.46&   8.09&22.7\\     
54&SPL &4.02&   4.87&$-$\\      
55&SPL &2.69&   3.31&$-$\\      
56&APL &0.52&   3.88&$-$\\
57&SPL &1.71&   2.11&10.8\\     
58&PL  &0.95&   6.38&$-$\\      
59&APL &36.0&   76.0&1.6\\      
60&SPL &10.64&  16.55&2.2\\     
61&SPL &7.07&   4.08&7.6\\      
62&SPL &8.97&   9.82&5.1\\
63&PL  &1.18&   7.21&5.4\\
64&SPL &6.29&   4.13&1.2\\
65&APL &0.12&   3.98&1.7\\
66&SPL &3.63&   4.36&27.1\\
67&SPL &2.00&   3.40&11.5\\ 
68&SPL &5.92&   13.41&3.6\\ 
69&APL &0.04&   4.20&$-$\\       
70&SPL &2.49&   3.63&5.9\\              
71&SPL &3.52&   5.57&4.9\\      
72&PL  &0.23&   6.76&$-$\\      
73&SPL &1.19&   1.14&$-$\\ 
74&APL &0.07&   1.55&$-$\\      
75&SPL &1.73&   1.78&$-$\\      
76&APL &0.50&   1.18&$-$\\                      
77&PL  &0.74&   3.55&$-$\\      
78&SPL &1.51&   1.51&$-$\\              
79&TM  &1.07&   1.03&$-$\\              
80&TM  &0.53&   1.81&$-$\\              
81&SPL &1.43&   1.15&$-$\\              
82 &SPL &2.64&  3.17&$-$\\      
83 &SPL &1.17&  1.32&$-$\\      
84 &APL &0.38&  2.16&$-$\\      
85 &SPL &5.12&  2.49&5.8\\      
86 &SPL &3.04&  4.41&$-$\\      
87 &SPL &3.27&  2.76&$-$\\      
88 &APL &0.58&  1.73&$-$\\              
89 &APL &1.92&  5.07&$-$\\      
90 &SPL &1.96&  1.65&$-$\\      
\hline
\end{tabular}
\end{center}
$^{(a)}$ in units of 10$^{-14}$ \cgs~ $^{(b)}$ in units of 10$^{44}$ erg s$^{-1}$\\
\end{table}
The results inferred by the spectral analysis of the entire sample can be briefly summarized as follows.
For about 65\% (i.e. 60 out of 90) of the X--ray sources the SPL model represents an acceptable description of their spectra.
26 (out of 90) sources require the introduction of a significant absorption component 
(see Table~\ref{tab:APL}, Table~\ref{tab:complex} and \p1). The resulting \nh~distribution is shown in Figure~\ref{fig:absorption}.
Furthermore, 13 sources require a more complex fitting model than SPL or APL to account for a soft excess component
(11 out of 13) or for the presence of warm absorber signatures (2 out of 13).
  
Measured values of the hard X--ray flux range  from $\sim$ 1 to $\sim$ 80 
$\times$ 10$^{-14}$ \cgs, with more than 50 out of 90 sources (i.e. 55\%) at \fhx~\simlt~5$\times$ 10$^{-14}$ \cgs~ i.e. flux levels
almost unexplored by the X--ray telescopes operating before \xmm. 
As expected on the basis of our selection criterion, in the soft X--ray band we detect sources in a broader flux range i.e.
from  \fsx~$\approx$ 70 down to $\approx$ 0.04 $\times$ 10$^{-14}$ \cgs.
The absorption--corrected 2--10 keV luminosities span from $\approx$ 2 $\times$ 10$^{40}$ \ergs~to $\approx$ 5 $\times$ 
10$^{45}$ \ergs~ in agreement with the optical classification of the identified sources in the sample. 
All but one (i.e. NGC 4291, n.~37) sources have a \lum~\simgt~ 10$^{42}$ \ergs~ typical of AGN: the two optically ``dull'' galaxies, i.e. source Nos. 3 and 41 (see \p1), too.\\ 

Before  drawing conclusions from the results of the X--ray spectral analysis,
we have checked out the possible presence of systematic trends due to the source position in the detector plane  
which could affect photon index and/or X-ray flux measurements.
Fig.~\ref{fig:gammahard} shows the photon index obtained with the SPL 
model plotted against the hard X--ray flux: many sources have a $\Gamma$ $\sim$ 1.8--2 i.e. the typical value
of unabsorbed AGNs (George et al. 2000; Nandra \& Pounds 1994).
\begin{figure}[b]
\begin{center}
\psfig{file=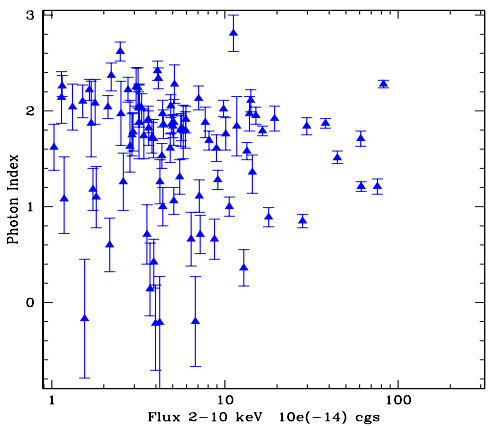,height=6cm,width=7cm}
\caption{Photon index obtained with the SPL model as a function of the flux in the 2--10 keV band.}
\label{fig:gammahard}
\end{center}
\end{figure}
This matches well with the optical identifications as well
as with the results obtained in  the {\it ASCA} Large Sky Survey (Akiyama et al. 2000).
On one hand, this plot shows that sources with flat slopes ($\Gamma \leq$ 1.3)
are present at various flux levels and no trend of $\Gamma$ versus flux is evident in the data.
On the other hand, despite the low number of objects at \fhx~\simgt~\f13~\cgs, very flat (i.e. $\Gamma$ \simlt 0.6) 
and inverted spectrum sources appear to be located in the region of the lowest fluxes as predicted by the CXB synthesis 
model and recently observed in \chandra~deep surveys (Tozzi et al. 2001; Stern et al. 2002).

We have calculated the average SPL spectral indices in the 0.3--10 keV band for the BRIGHT and the FAINT subsamples 
(see Sect.~\ref{lab:obs}) and obtained $\langle\Gamma\rangle$ = 1.54 $\pm$ 0.03 and  $\langle\Gamma\rangle$ = 1.56 $\pm$ 0.05, respectively. 
These results and their comparison with the values obtained
from other hard X--ray surveys will be extensively discussed in Section~\ref{lab:averageslope}.\\

\begin{figure}[t]
\begin{center}
\psfig{file=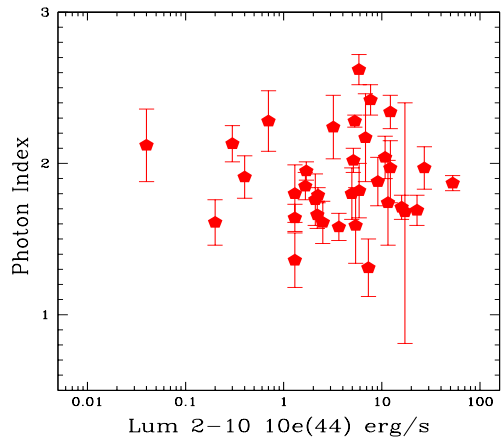,height=6cm,width=7cm}
\caption{Spectral index, computed assuming the best fit model for all optically identified sources in the sample (except for NGC 4291), 
as a function of hard X--ray absorption--corrected luminosity.}
\label{fig:gamma_lum}
\end{center}
\end{figure}
\begin{figure}[t]
\begin{center}
\psfig{file=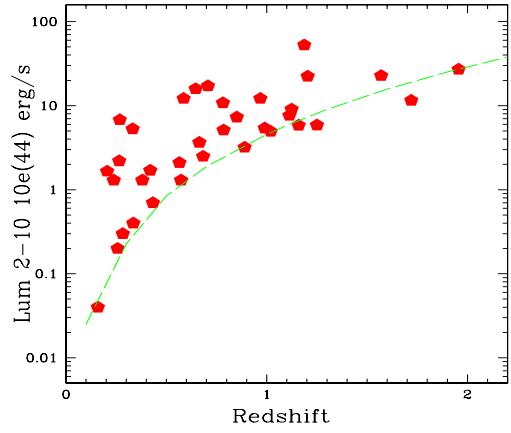,height=6cm,width=7cm}
\caption{Intrinsic 2--10 keV X--ray luminosity plotted against redshift for the identified sources
in the sample (except for NGC 4291). The dashed line indicates the ``completeness'' 
flux limit of our survey ($\sim$ 5 $\times$ 10$^{-14}$
 \cgs) once all the FOVs are taken into account.}
\label{fig:lum_z}
\end{center}
\end{figure}
In Fig.~\ref{fig:gamma_lum} we plot the photon index  as a function of the luminosity 
in the 2--10 keV band, as obtained applying the best fit model for each optically 
identified source in the sample.
There is no apparent trend for spectral variations as a function of luminosity. However, a large dispersion in the slope
values is present: this could either be due to intrinsic differences (George et al. 2000) or to the contribution from additional spectral
 components (i.e. reflection, soft--excess), unresolved here due to the limited statistics.  
  
Fig.~\ref{fig:lum_z} shows the luminosity--redshift relation for the optically identified sources in the sample:
as expected, the minimum and the maximum hard X--ray luminosity in each redshift bin 
increases with the redshift owing to a selection effect (most luminous sources are detected farthest). 
The present dataset is too small to disentangle any true luminosity
dependence (as claimed in Barger et al. 2002) from any mere selection effects. 
\subsection{Multiwavelength properties of the sources}
A classical approach (Maccacaro et al. 1988) extensively used in the X-ray surveys to infer some 
information about the nature of unidentified sources as well as to test the reliability of the 
optical identifications themselves is the so--called ``$F_X$/$F_{Opt}$'' diagnostic diagram. 
Since $R$ magnitudes are available for the majority of the sources in the sample (Table~\ref{tab:sample}), 
we adopt this band to build such a diagnostic diagram.
The relation between optical magnitude and hard X--ray flux
is plotted in Fig.~\ref{fig:multiband}.
The flux in the $R$ band is related to the $R$ magnitude by the following formula: 
$log F_R =-5.5 -0.4 \times R$ (Hornschemeier et al. 2001); hence, the relationship 
between optical flux and X--ray flux, obtained by Maccacaro et al. (1988) for
the $V$ magnitude, becomes the following: $log (F_X/F_R) = log F_X +5.5 + R/2.5$.

The three dashed lines  in  Fig.~\ref{fig:multiband} represent the X--ray to optical flux ratio
 of  $log (F_X/F_R)$ = 1, $-$1 and $-$2 (from top to bottom, respectively).
They mark the regions occupied by different classes of X--ray sources as indicated in the Figure.
A value comprised between $-$1 $< log (F_X/F_R) <$ 1 is typical of the ``standard'' luminous AGNs, 
both of Type 1 and Type 2. 
As expected on the basis of the
optical identifications, most of the identified sources (indicated with {\it triangles} in the Figure) are properly
located in the AGN locus. Interestingly, also most of the unidentified sources (indicated with {\it pentagons} in the Figure)
fall within the same region. 
This fact strengthens the results of our X--ray spectral analysis which indicated typical AGN--like spectra for all of them.
These findings fully agree with results from hard X--ray surveys which found AGNs as the dominant population
(e.g. Akiyama et al. 2000; Barger et al. 2002; Hasinger 2003).

However, there are a few sources that have values of $log (F_X/F_R)$ outside the typical AGN locus. In particular,  
there are 4 objects (i.e. Nos. 25, 38, 63 and 65) that are notably faint in the optical with respect to their X--ray
 flux values. They are of particular interest because results from \chandra~and
\xmm~deep surveys (e.g. Alexander et al. 2001; Mainieri et al. 2002) suggest that this kind of
sources are a mixture of ``exotic'' AGNs such as Type 2 QSOs, Extremely Red Objects (EROs, with $R-K$ \simgt 5), high 
redshift ($z \geq$ 3) galaxies which host a dust--enshrouded AGN and, possibly, QSOs at  $z$ \simgt~ 6.
\begin{figure}[!]
\begin{center}
\psfig{file=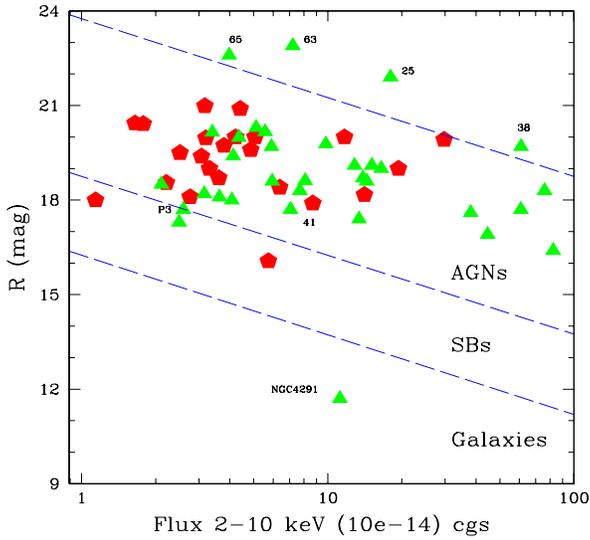,width=8cm,height=7.5cm}
\caption{Plot of optical $R$ magnitude versus hard X--ray flux for those sources in the sample with available $R$ band photometry.
The optically identified (unidentified) sources are indicated with {\it triangles} ({\it pentagons}).
The dashed diagonal lines have been calculated following Hornschemeier et al. (2001) and 
represent the constant flux ratios $log (F_X/F_R) =$ 1 (upper), $-$1 and $-$2 (lower). Labels show the
typical  $(F_X/F_R)$ range observed for luminous AGN, starburst galaxies and normal galaxies in the local Universe.
Peculiar sources (see text) are also indicated.}
\label{fig:multiband}
\end{center}
\end{figure}

All the sources in the sample with  $log(F_X/F_R) >$ 1, i.e. optically--weak, are found indeed significantly X--ray absorbed as shown in 
Tables~\ref{tab:APL} and~\ref{tab:complex}, and, interestingly, they are also all RL QSOs: two of these are 
also classified as narrow line QSOs of the class of EROs (i.e. Nos. 63 and 65), and
the remaining two sources (i.e. Nos. 25 and 38) are optically classified as broad line QSOs. These latter show moderately
flat X--ray spectra with best--fit photon indices $\Gamma \simeq$ 1.5--1.6 which are typically observed in core--dominated 
RL QSOs (Shastri et al. 1993; Sambruna et al. 1999; Reeves \& Turner 2000). Photometric information in the $K$ band
could be particularly useful to understand whether also these two QSOs could be classified as EROs as well.

At the fluxes currently sampled by our survey, the population of objects with low (\simlt~--1) and very low 
($<$ --2) values of $log (F_X/F_R)$ are largely missed. Such kind of sources has been found to slowly
emerge just below \fhx~$\approx$ 10$^{-15}$ \cgs~and are thought to be mainly star--forming galaxies, LINERs and normal galaxies 
(Barger et al. 2002).
These classes of sources are commonly detected in the soft X--ray band, where they are expected to be the
dominat population at \fsx \simlt  10$^{-17}$ \cgs~(Miyaji \& Griffiths 2002). 
Three sources (Nos. 14, 73 and 85) in our sample show instead a ratio $log(F_X/F_R) <$  $-$1.
Among them, the first two are optically unidentified while the latter is a broad line QSO. The X--ray spectrum
of this faint quasar (\fhx~$\approx$ 2.5 $\times$ 10$^{-14}$ \cgs) is very steep ($\Gamma \sim$ 2.6, e.g. Table~\ref{tab:SPL}), 
with the source counts fairly dominated by the soft X--ray photons. This is probably the reason why we measure a value 
of $log (F_X/F_R)$ slightly lower than unity. 

As expected, the normal galaxy NGC 4291 (i.e. No. 37) is the only object with a $log (F_X/F_R) <$ $-$2. 
Besides normal galaxies, also nearby Compton--thick AGNs usually exhibit such a low $log (F_X/F_R)$ value (Comastri et al. 2003).
Accordingly, it is worth noting that the peculiar sources No.~3 (the ``P3'' galaxy, e.g. Fiore et al. 2000) 
and No.~41 (that are optically identified as ``normal'' galaxies, see \p1) are found to lie in the region of the diagram typical of AGNs. 
This finding further supports the obscured AGN nature of these two objects.

\section{Hard X-ray selected QSOs}
\label{HXSQs} 

A meaningful product of our wide--angle survey consists in the opportunity 
of investigating  at faint flux levels
the spectral properties of hard X--ray selected quasars (hereafter HXSQs) and of 
comparing them with those derived from other studies of soft X--ray and/or optically selected QSOs.

\begin{table}[b]
\caption{Properties of hard X--ray selected QSOs.}
\label{HXQSOs}
\begin{center}
\begin{tabular}{ccccc}
\hline
\multicolumn{1}{c} {Objects}&
\multicolumn{1}{c} {N.}&
\multicolumn{1}{c} {$\langle\Gamma_{Best-fit}\rangle$} &
\multicolumn{1}{c} {$\langle\Gamma_{SPL}\rangle$} &
\multicolumn{1}{c} {$\langle L_{2-10}\rangle$$^{a}$}\\
\hline
\hline
         &                &             &\\
All QSOs &30&1.87$\pm$0.04&1.66$\pm$0.03&8.97\\  
QSOs at $z<$1     &20&1.83$\pm$0.05&1.52$\pm$0.04&5.96\\  
QSOs at $z\geq$1&10&1.96$\pm$0.05$^{b}$&1.96$\pm$0.05$^{b}$&14.98\\
\hline 
\end{tabular} 
\end{center}
$^a$ Luminosity in units of 10$^{44}$ erg s$^{-1}$.\\
$^b$ Excluding source No.85: $\langle\Gamma_{SPL}\rangle$$\equiv$$\langle\Gamma_{Best-fit}\rangle$=1.88$\pm$0.05.
\end{table}
In order to extract QSOs from the list of  sources with optical identification, 
we select those showing an intrinsic 2--10 keV luminosity larger 
than 10$^{44}$ erg/s. 
In doing so, we create a sample of 30 HXSQs that is one of the largest sample of this kind available to date.
Their redshifts range from $z$ = 0.205 to $z$ = 1.956 with a mean  $\langle z \rangle$ $\sim$ 0.7;
the vast majority of the objetcs are broad lines AGNs with the remarkable exceptions of source Nos.~63 and 65, which show 
 an optical Type 2 classification (Table~\ref{tab:sample}; Mainieri et al. 2002).    
The fraction of these HXSQs that turn out to be radio loud (i.e. with $\alpha_{OR} \geq$~0.3) is 20\% (i.e. 6 out of 30 sources) but,   
keeping in mind the incompleteness affecting the optical and radio coverage of our sample (see Table~\ref{tab:sample}), 
this fraction should be considered only as a  lower limit.
Furthermore, we are able to split the sample of HXSQs in two subsamples having 10 and 20 objects 
each and redshift 
larger and lower than 1, respectively. 
The average properties drawn from the spectral analysis of these two samples of HXSQs (total and subsamples) are reported in 
Table~\ref{HXQSOs}, while the photon indices obtained with the SPL and with the best fit model found in each individual HXSQ 
are displayed in Fig.~\ref{fig:QSOevol}~and Fig.~\ref{fig:QSObest}, respectively.

When the SPL model is applied we find a $\langle\Gamma\rangle$ = 1.66 $\pm$0.03 which rises to $\langle\Gamma\rangle$ =
 1.87$\pm$0.04 using the best fit model instead. This effect can be easily explained by the presence of intrinsic absorption which suppresses the soft
 portion of the X--ray emission. In particular, only HXSQs at $z <$ 1,
 either radio--quiet (RQ) or radio--loud (RL), show this excess of absorption. 
Because of the limited number of HXSQs with $z >$ 1, the corresponding value of  $\langle\Gamma\rangle$ 
appears to be biased by the very steep index of LBQS~2212--1759 (source No. 85).
This source yields indeed a $\Gamma \sim$ 2.6, similar to what found by  Brinkmann et al. (2003) in another bright RQ QSO: 
such a steep slope is reminescent of  Narrow line Seyfert 1 galaxies, which usually show a strong soft excess
 component. However, any such soft excess component would be redshifted almost out of the \epic~energy range given the source
redshift of $z$ = 1.159. In any case, a more detailed modeling of such 
soft excess component is not allowed due to the low photon statistics. 
Excluding source No. 85 from the sample, we obtain indeed an average index of $\langle\Gamma\rangle$ = 1.88$\pm$0.05,
consistent with the average index found for the HXSQs at $z <$ 1. 
We can therefore conclude that the average X--ray spectral slope of HXSQs resulting from the present work 
is almost the same from $z \approx$ 0 to $z \approx$ 2 indicating no spectral evolution, i.e. 
no obvious variation of $\Gamma$ along the redshift (see Fig.~\ref{fig:QSObest}). 

\begin{figure}[th]
\begin{center}
\psfig{file=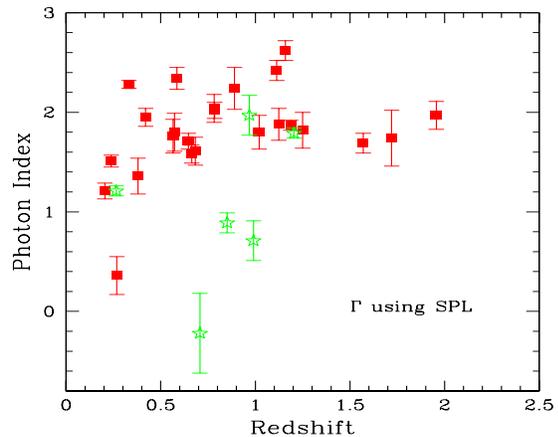,width=7.5cm,height=6cm,angle=0}
\caption{Plot of the full--band (0.3--10 keV) spectral index vs. redshift using SPL model  for all the hard X--ray selected QSOs (HXSQs).
 The filled squares and the void stars are the ``radio--quiet'' HXSQs and the ``radio--loud'' HXSQs, respectively. 
It is worth stressing the incomplete radio coverage of our X-ray sample, so we include in  ``radio--quiet'' HXSQs
 also some objects with unknown radio properties.}
\label{fig:QSOevol}
\end{center}
\end{figure}
\begin{figure}[htb]
\begin{center}
\psfig{file=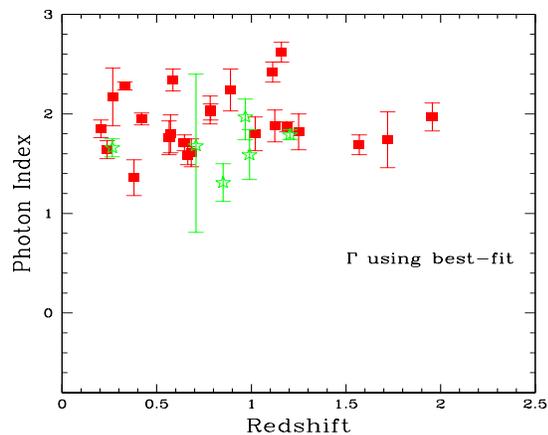,width=7.5cm,height=6cm,angle=0}
\caption{Same as in Figure~\ref{fig:QSOevol}, but with spectral indices derived using the best fit model for each source.}
\label{fig:QSObest}
\end{center}
\end{figure}

\subsection{Comparison with previous works}
The value of $\Gamma \approx$ 1.8--1.9 is similar to the mean slope of unobscured local Seyfert-like AGNs
 (i.e. $\Gamma$ = 1.86$\pm$0.05, Nandra et al. 1997; Perola et al. 2002; Malizia et al. 2003) and it matches well with previous
 X--ray studies of quasars carried out with different telescopes (Comastri et al. 1992; Lawson \& Turner 1997).
 In particular, Reeves \& Turner (2000) analysed a sample of 62  QSOs with {\it ASCA}, covering a redshift range
 from 0.06 to 4.3, and they also reported a comparable mean photon index $\langle\Gamma\rangle$ = 1.76$\pm$0.04. 
They also claimed a difference between the slopes of RQ QSOs ($\Gamma \sim$ 1.9) and RL QSOs ($\Gamma \sim$ 1.6),
 with the latter showing a flatter average photon index. This issue has been extensively discussed in Sambruna,
 Eracleous \& Mushotzky (1999, hereafter SEM99): these authors concluded that since lobe--dominated RL QSOs have
 the same intrinsic photon indices of RQ QSOs, the observed harder X--ray spectra of core--dominated RL QSOs being
 likely due to the presence of a beamed  extra--continuum component originating in the inner parts of the radio jet.
For the six RL HXSQs in our sample we find a  $\langle\Gamma\rangle$ = 1.66$\pm$0.17 which is slightly harder but still consistent
 with the typical value measured for RQ QSOs. However, it is worth noting that source Nos.~25 and 38, two bright objects
 with high--counting statistics, show flat photon indices ($\Gamma \sim $ 1.3 and 1.5, respectively) similar to those usually
 found in core--dominated RL 
QSOs; however in these two sources we also find that the flat spectrum is due to the presence of intrinsic absorption.

Our data also confirm that RL QSOs are characterized by absorption in excess to the Galactic value
 (Elvis et al. 1994): 4 out of our 6 RL objects are indeed obscured by column densities of $\approx$ 10$^{21-23}$ cm$^{-2}$.
 It has been suggested in other works (i.e. Cappi et al. 1997; SEM99), also on the basis of variability 
arguments, that such an absorber could be located in the inner nuclear regions of RL QSOs rather than associated with either
the host galaxy or a surrounding cluster. 
Moreover, similar to our finding, SEM99 found that \simgt 50\% of broad line RL QSOs in their sample suffered from significant
 intrinsic neutral absorption. These authors considered this 
fact at odds with simple orientation--based AGN unification models and  
suggested the possibility of different physical conditions of the gas around
the central engine in RL QSOs, being colder in these objects than
in RQ ones (George et al. 2000).

\begin{figure}[tb]
\begin{center}
\psfig{file=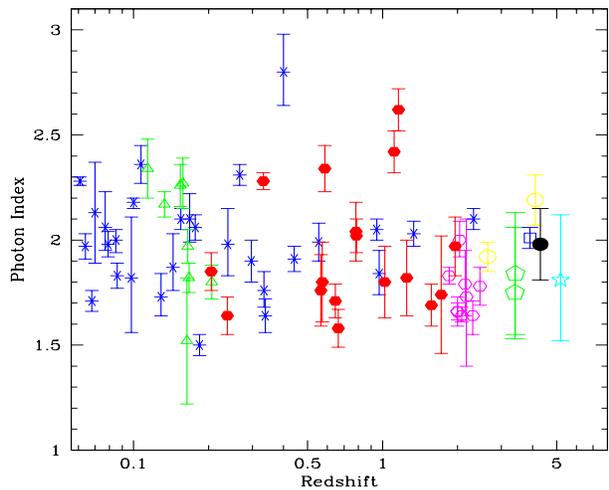,width=8.5cm,height=7cm,angle=0}
\caption{Photon index versus redshift for a collection of RQ QSOs observed with different X--ray telescopes.
Data points are taken, respectively, from: {\em ASCA}  ({\it asterisks}, Reeves \& Turner 2000 and George et al. 2000),
({\it void exagons}, Vignali et al. 1999); {\em BeppoSAX}  ({\it void triangles}, Mineo et al. 2000).
 Data from the present work are indicated with {\it filled exagons}: 
we have excluded the QSOs from the PKS 0312--770 field because of this field has no NVSS radio coverage.
 The {\it void pentagons} indicate XMMU J105144.6$+$572807 and XMMU J105125.3$+$573048 (both at $z \sim$ 3.4)
 which are the only RQ QSOs at $z >$ 3 from the \xmm~long--exposure of the Lockman Hole (Mainieri et al. 2002).
 The {\it void circles} represent two QSOs at $z$ = 2.64 and $z$ = 4.1 observed with \xmm~(Ferrero \& Brinkmann 2003).  
The other three points at the highest redshifts correspond, respectively, to: APM 08279$+$5222 ({\it void square};
 $z$ = 3.91, e.g. Hasinger et al. 2002),  a joint spectrum of 9 \chandra~QSOs with 4.09 \simlt $z$\simlt~ 4.51 plotted here
 at their average redshift ({\it filled circle}, Vignali et al. 2003) and CXO J123647.9$+$620941 ({\it void star}
; $z$ = 5.186, e.g. Vignali et al. 2002).}
\label{fig:tuttiQSO}
\end{center}
\end{figure}
\subsection{Spectral evolution}
Concerning the evolution of the mean spectral shape, the value assessed for HXSQs by our analysis provides further support to the 
recent findings of Vignali et al. (2003). In fact, these authors invoke a universal X--ray 
emission mechanism for QSOs, i.e. indipendent from cosmic time and luminosity, on the basis of the analysis of nine 
high redshift ($z >$ 4) QSOs 
observed with {\em Chandra}.
By a ``stacked'' spectral fitting, they found an average
spectral slope of 1.98 $\pm$ 0.16 in the 2--30 keV rest--frame band (but see also Bechtold et al. 2002).

We report in  Figure~\ref{fig:tuttiQSO}  
a large compilation from literature of photon indices obtained from the spectral analysis of radio--quiet QSOs  up to 
$z$ $\approx$ 5.2 observed with different X--ray telescopes that includes also the recent \chandra~results mentioned
 above together with those found in the present work. 
No evident trend of the spectral slope appears to emerge along with the redshift from this plot, confirming that the accretion 
mechanism in RQ QSOs is the same at any redshift sampled to date.

\section{Observational constraints for CXB synthesis models}
\label{cxb}
Two main observational constraints on the predictions of synthesis models can be extracted from the present work:
 (1) the average slope of the
sources making the CXB at intermediate flux levels and (2) the corresponding ratio of absorbed--to--unabsorbed sources. Both these 
parameters are essential in the definition of the average spectrum at a given flux level in order to estimate the relative 
contributions of individual classes of objects to the CXB.
\subsection{Average slope at faint fluxes}
\label{lab:averageslope}
The average photon index calculated over the 0.3--10 keV band with the SPL model  
is $\langle\Gamma\rangle$ = 1.59 $\pm$ 0.02. 
As expected, once the best fit model of each source is assumed\footnote{We exclude from this calculation NGC4291 (No.~37) 
as its spectrum is better described by a thermal model, in agreement with its optical classification as a normal galaxy (see \p1).}, 
the resulting value of the average slope  becomes steeper i.e. with a  $\langle\Gamma\rangle$ =  1.80 $\pm$ 0.04.
This value agrees with the typical one of unabsorbed AGNs found in previous works (e.g. Lawson \& Turner 1997; Reeves \& Turner 2000; 
Malizia et al. 2003) and it is in agreement with the classification as broad line AGNs for most of the identified sources
(see Table~\ref{tab:sample}).\\

By dividing the whole sample
into the BRIGHT and the FAINT subsamples, both of which are complete at their flux limit (\fhx~\simgt~5 $\times$ 10$^{-14}$ \cgs~and
\fhx~\simgt~2 $\times$ 10$^{-14}$ \cgs) we calculate in each case the average slope with the SPL model 
which turns out to be $\langle\Gamma\rangle$ = 1.54 $\pm$ 0.03 and  $\langle\Gamma\rangle$ = 1.56 $\pm$ 0.05, respectively. 
\begin{figure}[t]
\begin{center}
\psfig{file=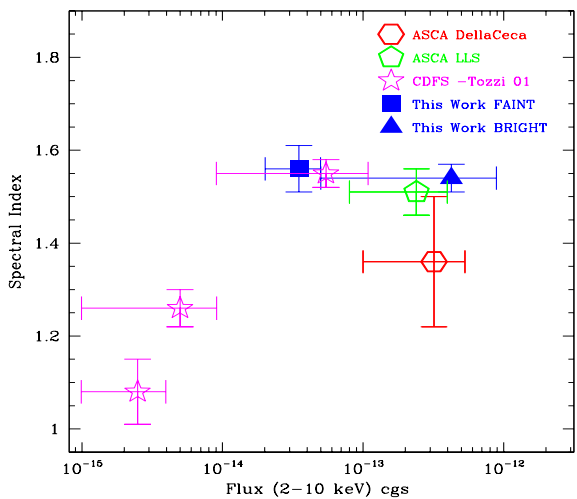,width=8cm,height=7cm,angle=0}
\caption{Average spectral index from a power model model (SPL) as a function of the hard X--ray flux. Data are taken from:
Della Ceca et al. (1999; {\it exagon}), Ueda et al. (1999; {\it pentagon}), Tozzi et al. (2001; {\it stars}). The {\it triangle} 
({\it square}) represents the value obtained in the present work for the BRIGHT (FAINT) subsample.}
\label{fig:index}
\end{center}
\end{figure}

In Fig.~\ref{fig:index} the average photon indices of the BRIGHT and FAINT samples are compared with those obtained by
the most popular hard X-ray surveys carried out so far. 
It is worth noting that our data resulted from the spectral analysis of individual sources,
whereas all other values were derived by the stacking technique. A good agreement between our values and those
obtained in other works is evident from this plot. 

At \fhx~\simgt1 $\times$ 10$^{-14}$ \cgs~the average slope appears
still steeper than the integrated spectrum of the CXB, which has $\Gamma$ = 1.4, thus suggesting that the bulk of
 the flat spectrum (i.e. absorbed) sources
 has still to emerge at these flux levels. The measurements obtained by the \chandra~Deep Field South survey
 (Tozzi et al. 2001; indicated as {\it stars} in Fig.~\ref{fig:index})
confirm indeed this suggestion, revealing a progressive and significative flattening of the average spectral
 index below  \fhx~\simgt~10 $\times$ 
10$^{-14}$ \cgs, which is able to solve the ``spectral paradox''.\\

As mentioned in Sect.~\ref{intro}, optical follow--up observations of X--ray deep surveys 
have recently pointed out that the bulk of the hard CXB is produced by a large number 
of narrow line active galaxies at $z \sim$ 0.7 having Seyfert--like luminosities. 
Consequently, we have performed a test to show how the spectral hardening occuring at \fhx~$\sim$ 10$^{-15}$ \cgs~
could be qualitatively explained by these sources.
Accordingly, we have assumed a typical X-ray spectral template of a Seyfert 2 galaxy as suggested in 
Turner et al. (1997), i.e. a  partial covering model with $\Gamma$ = 1.9, \nh~= 10$^{23}$ \cm2~and a hard X--ray (deabsorbed) luminosity of 10$^{43}$ \ergs. We have further assumed $z$ = 0.7 for the redshift value. 
The flux value calculated with these assumptions results to be equal to \fhx~$\approx$ 4 $\times$ 10$^{-15}$ \cgs, i.e. well 
in the range where the spectral flattening has indeed been observed (Fig.~\ref{fig:index}). 
In addition, it is worth stressing that our simulations\footnote{Assuming an input Seyfert 2--like spectrum as
reported above and a 300 ks \pn~exposure, we found a photon index $\Gamma$ = 0.92$_{-0.43}^{+0.51}$ in the case of a 
fit with the SPL model.} have shown that
such a source would show a flat
continuum if it  were fitted with SPL model, i.e. as observed in an  heavily absorbed object.

If the bulk of the hard CXB originates from a large population of Compton--thin 
low--redshift Seyfert 2--like  galaxies, they could indeed account for the
flattening of the average photon index observed by \chandra~towards very faint flux levels. 
\subsection{Fraction of absorbed sources}
\label{lab:fraction}
Fig.~\ref{fig:fraction} shows the fraction of absorbed (i.e. with an \nh~\simgt~10$^{22}$ \cm2~ in excess to the Galactic value)
 to unabsorbed sources that we find in the BRIGHT and FAINT subsamples together with those obtained by other
 hard X--ray surveys with different flux limits in the flux range
from \fhx~$\sim$ 10$^{-11}$ down to \fhx~$\sim$ 10$^{-14}$ \cgs. 

Interestingly, it appears from this Figure that the fraction of absorbed sources remains almost the same ($\sim$ 30\%) 
in this wide range of hard X--ray fluxes. 

It is worth noting that in order to calculate the fractions for both samples
reported in this Figure we assume $\Gamma$ = 1.9 for all the sources and $z$ = 1 for the unidentified ones 
to overcome some biases which may affect our estimates, as mentioned in Sect.~\ref{basicmodels}. 
The values reported in Fig.~\ref{fig:fraction} can be therefore considered as  conservative estimates of the fraction of absorbed sources in both samples. 
\subsubsection{Comparison with theoretical predictions}

This finding is fairly unexpected since all synthesis models of the CXB 
(e.g. Gilli et al. 2001; Wilman \& Fabian 1999; Comastri et al. 2001, hereafter C01) predict that as fainter fluxes
 are considered, more absorbed sources should be found.\\
In particular,  in Figure~\ref{fig:xrb_mybin} the theoretical predictions of the CXB synthesis model of C01 are plotted together
 with the fraction of absorbed sources in the 2--10 keV band found in our spectral survey.
\begin{figure}[t]
\begin{center}
\psfig{file=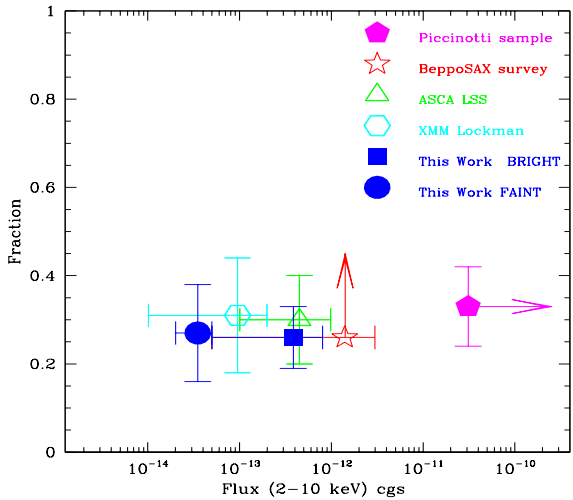,width=8cm,height=7cm,angle=0}
\caption{Fraction of absorbed sources (i.e. with \nh ~\simgt 10$^{22}$ cm$^{-2}$) taken from different hard X--ray surveys.
 The {\it filled square} and the {\it filled circle} indicate the values found from our BRIGHT (60\% optically identified, 
hereafter ID) and FAINT ($\sim$ 35\% ID) samples, respectively.
Other symbols represent the values from: Piccinotti sample (Comastri 2000, but also including 4 BL Lacs and M82 present
in the original sample of Piccinotti et al. 1982; {\it filled pentagon}; 100\% ID), {\em BeppoSAX} 2--10 keV survey (Giommi et
al. 2000); {\it void star}), {\em ASCA} Large Area Survey (Akiyama et al. 2000; {\it void triangle}; 97\% ID), and \xmm ~survey in 
the Lockman Hole (Mainieri et al. 2002, {\it void exagon}; $\sim$ 90\% ID).
The horizontal error bars associated to the points represent the flux interval between  
the completeness limit and the highest flux sampled in each survey; 
the vertical error bars indicate instead the Poissonian errors for each fractional value.} 
\label{fig:fraction} 
\end{center}
\end{figure}
For a better comparison we have splitted our measurement regarding the BRIGHT sample into three points.
Moreover we have added the point relative to 9 $\times$ 10$^{-15}$ \simlt~\fhx~\simlt~ 2 $\times$ 10$^{-14}$ \cgs~calculated
 taking into account all
the sources selected in the field of LBQS 2212--1759 (this field has indeed the longest exposure time amongst the observations
 considered here, i.e. 80.5 ks, see Table~\ref{tab1}) and those in the 100 ks exposure of the Lockman Hole presented in Mainieri et al. (2002).
\begin{figure}[t]
\begin{center}
\psfig{file=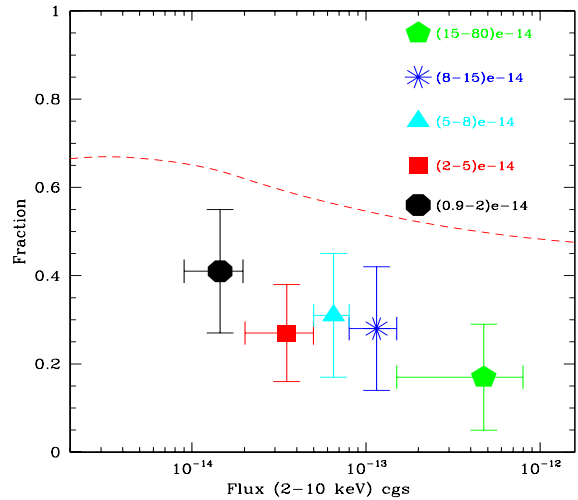,width=8cm,height=7cm,angle=0}
\caption{Fraction of absorbed sources  (i.e. with \nh ~\simgt 10$^{22}$ cm$^{-2}$) from our analysis vs. 2--10 keV flux
 compared to theoretical prediction of Comastri et al. (2001). For the calculation of our points we assumed a conservative
 spectral slope of $\Gamma$=1.9 for all sources and $z$ = 1 whenever the X--ray source was not optically identified.
 The value for the point at faintest flux has been calculated using sources selected in the deepest exposure of ours
 (i.e. the LBQS~2212$-$1759 filed, 80.5 ks) together with  sources in the 100 ks exposure of the 
Lockman Hole analysed by Mainieri et al. (2002).}
\label{fig:xrb_mybin}
\end{center}
\end{figure}

This figure shows that the fraction of absorbed sources predicted by the theory is a factor of $\sim$ 2 larger (with a $\approx$ 4.5$\sigma$ 
significance) than our observational data at comparable fluxes.
We thus confirm and extend even to fainter fluxes the data--to--model mismatch found in \p1. 
Our data suggest that  unabsorbed (i.e.with \nh~\simlt~10$^{22}$ \cm2) objects still largely dominate the source counts 
at  $F_{2-10}$ \simgt 10$^{-14}$ \cgs, at odds with the theoretical expectations of about 50--60\% of absorbed sources.

Despite this mismatch, the overall trend predicted by the model follows the observational data as if only the normalization was wrong.

Although very preliminary, also the results of the individual 
spectral analysis of the brightest X--ray sources in the HDFN (Bauer et al. 2003) seem to suggest
a \nh~distribution peaked towards low column density values with only few sources ($<$ 9\%) obscured by \nh~\simgt~10$^{23}$
\cm2~whereas the synthesis models predict a value of $\sim$ 20\%.   
Other works based on hardness ratios analyses further confirm such a scarsity of obscured objects above $F_{2-10}$ $\sim$ 10$^{-14}$ \cgs~
(e.g. Baldi et al. 2002; Akiyama, Ueda \& Otha 2002): in particular the latter authors reported a fraction of absorbed source with   
 \nh~\simgt~10$^{22}$ \cm2~and \lum~$>$ 10$^{44}$ \ergs~of $\sim$ 15\%, i.e. a factor of 3 lower than expected. 

Furthermore, the average photon index values at \fhx~\simgt~10$^{-14}$ \ergs~reported in the present as well as in other works 
(Sect.~\ref{lab:averageslope}) result to be steeper than the slope of the integrated CXB spectrum:
this result is indicative of the fact that absorbed objects are not yet present in large quantities at these fluxes, as 
suggested instead by theoretical models.

Possible explanations for this observational mismatch will be extensively discussed in Sect.~\ref{lab:how}.

\subsection{Comparison with the results from the 1Ms HDFN survey}
\label{lab:hdfn}
Since this data/model mismatch on the fraction of absorbed sources has been claimed in \p1~and
in the present paper for the first time, it requires further investigations before validation. 
We have thus used the data from the \chandra~1Ms Hubble Deep Field North survey 
published in Brandt et al. (2001) to estimate the fraction of absorbed versus unabsorbed objects at our flux levels and 
lower.

To this aim,
from the entire X--ray source catalog only the sources detected in the 2--8 keV band (265) have been selected.  
Their fluxes span a very large range i.e. 1.4 $\times$ 10$^{-16}$ \simlt~$F_{2-8}$~\simlt~ 1.2 $\times$ 10$^{-13}$ \cgs.
Of these 265 sources, Barger et al. (2002) presented the optical identification for 118 (i.e. $\sim$ 45\%).

Since Brandt et al. (2001) used the source fluxes corrected for vignetting, 
we calculate source--by--source  
the ``flux ratio'' $FR$ rather than the commonly--used ``hardness ratio''. This ratio is defined as follows:\\
\centerline{$FR$ = ($F_{2-8}$ -- \fsx)/($F_{2-8}$ $+$ \fsx)}\\
Similarly to the  ``hardness ratio'', this quantity is indicative of the ``flatness'' of an X--ray spectrum.
Accordingly, sources with $FR$ = 1 represent those without a positive detection in the soft X--ray band but seen in the hard X--ray band.
We then compare the $FR$ value of every source with that (hereafter indicated with $FR_{z}$) 
obtained\footnote{These values of $FR_{z}$ are determined using the A02 version of {\it PIMMS} (Mukai 2000).
For example, for a source  with  $\Gamma$ = 1.7, \nh~= 10$^{22}$ cm$^{-2}$ and $z$ =1,   the  corresponding $FR_{z}$ value is 0.41} for
an object with $\Gamma$ = 1.7, \nh~= 10$^{22}$ cm$^{-2}$ placed at the redshift of that source (or $z$ = 1
for an optically unidentified one). Accordingly, a source that shows $FR >$ $FR_{z}$ 
is X--ray obscured with an \nh~$>$ 10$^{22}$ cm$^{-2}$. 

Following this procedure, it is possible to estimate the fraction of absorbed sources in the \chandra~HDFN 1Ms exposure
at different hard X--ray flux levels. These results are displayed in Figure~\ref{fig:hdfn} together 
with those derived from our analysis.

\begin{figure}[!t]
\begin{center}
\psfig{file=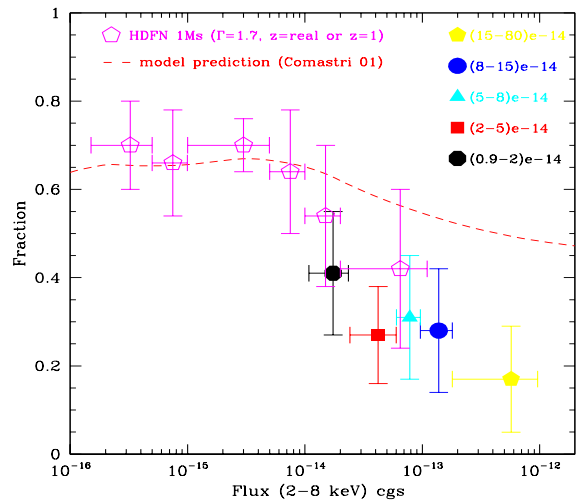,width=8cm,height=7cm,angle=0}
\caption{Ratios of absorbed (\nh~\simgt~10$^{22}$ cm$^{-2}$) to unabsorbed hard X--ray selected 
sources in our sample ({\it filled symbols}) and in the HDFN 1Ms exposures ({\it void pentagons}) as 
a function of the 2--8 keV flux  compared to model predictions ({\it dashed line}; Comastri et al. 2001). 
Points from our analysis are the same as those plotted in Fig.~\ref{fig:xrb_mybin},
but at flux values smaller than a factor of 1.2 (\fhx/$F_{2-8} \sim$ 1.2 with $\Gamma$ = 1.7) in order to provide a better 
comparison with HDFN data.
The values from the HDFN has been derived taking into account X--ray and optical data from
Brandt et al. (2001) and Barger et al. (2002), respectively. We obtained the fractions of absorbed sources in 
each flux bin by means the ``flux--ratio'', i.e.: $FR$ =($F_{2-8}-$\fsx)$/$($F_{2-8}+$\fsx), analysis.
We consider ``absorbed'' a source with an $FR$ value larger than that calculated assuming $\Gamma$ = 1.7, \nh~= 10$^{22}$ cm$^{-2}$ 
and the redshift proper of the source or  $z$ = 1 if the X--ray source is not optically identified 
(see text for further details).} 
\label{fig:hdfn}
\end{center}
\end{figure}

It appears from Figure~\ref{fig:hdfn} that at hard X--ray fluxes from $\approx$ 10$^{-14}$ \cgs~down to $\approx$ 
10$^{-16}$ \cgs~ the fraction calculated in the HDFN data matches very well with the model predictions. 
On the other hand, the two points at $F_{2-8} >$ 10$^{-14}$ \cgs~ are consistent both with the theoretical values and our estimates.
However, it is worth noting the large errors which affect 
the \chandra~measurements: in fact, due to their lower number densities per deg$^2$, the number of objects detected at the brightest
fluxes by this 1 Ms exposure is relatively small, with only 34 X--ray sources having
$F_{2-8} >$ 10$^{-14}$ \cgs. 
Furthermore, some flat spectra could be due to warm absorber features rather than cold absorption material 
and unfortunately hardness ratio analysis does not allow to discriminate between the two.

The analysis of this ultra--deep \chandra~observation therefore confirms only partially our finding.
Unfortunately, it does not provide
an efficient tool to disentangle the mismatch reported in Section~\ref{lab:fraction}.

This result is however very useful as it provides the most accurate estimate never published elsewhere
for the fraction of absorbed sources at flux levels much fainter than those sampled in our 
survey (\simlt  10$^{-15}$ \cgs).

\section{How to explain this data/model mismatch?}
\label{lab:how}
\subsection{Observational biases}
There are three main observational biases which may affect our estimate of the fraction of obscured sources, i.e.:
\begin{enumerate}
\item {\it Poor statistics from the faintest sources.} This fact may have lead in some cases to a 
possible underestimate of the real \nh~value because of the possible presence of an unresolved  soft--excess or reflection component.

However, we consider that this bias is in large part taken into account by fixing the photon index to $\Gamma$ =
1.9 before re--estimating the amount of cold absorption column density (see Sect.~\ref{basicmodels}).
Moreover, as reported in Sect. 4.2.2 of \p1, we performed simulations in order to rule out the possibility that we miss some obscured sources at fluxes above the completeness limit of our survey.

\item {\it Incompleteness of the optical identifications.} The column density measured in the optically
 unindentified sources would be certainly underestimated if they were placed at $z$ = 0 (see Sect.~\ref{basicmodels}).
 Nevertheless, we estimate that we overcome this bias by assuming $z$ = 1 
for such sources, i.e. by placing them just above the peak of the newly observed redshift distribution
 (e.g. Hasinger 2003; Barger et al. 2002) 
in order to provide a more conservative estimate of their \nh~values.

\item {\it Selection effect due to the PN effective area.}
The predictions shown in Fig.~\ref{fig:xrb_mybin} and Fig.~\ref{fig:hdfn} have been calculated assuming a flat--response detector (C01).
This is not the case of \xmm, whose effective area, although the best  to date,  is affected by
 a degradation  by about a factor of two between 4 and 9 keV.
Hence, a convolution of the models with the response of \epic~would be very important in order to allow a proper comparison between the
observational data and the theoretical expectations\footnote{Interestingly, Gilli et al. (2001) showed that the fraction of obscured AGNs
 (i.e. with \nh~$\geq$ 10$^{22}$ \cm2) predicted by their model at \fhx~$\approx$ 10$^{-13}$ \cgs~decreases by 20\%
 when folding the model predictions with the {\it ASCA SIS} effective area. 
If the fraction of obscured AGNs predicted by C01 is decreased by 20\% at  \fhx~$\approx$ 10$^{-13}$ \cgs~the discrepancy observed in 
Fig.~\ref{fig:xrb_mybin} would be consequently reduced.}. 

Interestingly, if we rescrict the analysis to the 5--10 keV energy band and compare the fraction of absorbed objects resulting from the complete 
subsample of 22 sources with $F_{5-10}$ \simgt~5 $\times$ 10$^{-14}$ \cgs~with the model prediction in this band, we still notice that our value
 is $\sim$ 30\%, i.e. a factor of 2 lower than  
expected by C01 ($\sim$ 60\%). 
\end{enumerate} 

\subsection{Revision of some model assumptions}
We address here the possibility 
that one (or more) of the assumptions usually included in synthesis models of the CXB need to be revised.
The results presented here and the mismatch emerging in the optical follow--up of the deep X-ray surveys
 (Hasinger 2003) between the predicted and the observed redshift distribution of X--ray sources indeed suggest that
some inputs of the CXB theoretical models should be updated.

The most critical assumptions usually made in theoretical works are three: the spectral template adopted for the X--ray sources,  
the distribution of absorption column densities and the slope, form and evolution of the X--ray luminosity function (XLF), which is 
almost unknown for Type 2 objects.

\begin{enumerate}
\item {\it The X--ray spectral properties.} Although the overall X--ray spectral shape of AGN is thought to be known
 (Mushotzky et al. 1993, Nandra \& Pounds 1994), 
some details are still uncertain.

For example, the average value of the high--energy cutoff (due to thermal Comptonization) is poorly constrained. 
Most theoretical 
models assume $E_{cutoff}$ = 320 keV (C01, Gilli et al. 1999), but this is certainly a rough approximation.
On the basis of {\it BeppoSAX} observations Matt (2000), Perola et al. (2002) and, recently, Malizia et al. (2003) clearly showed 
that the average cutoff value is more around 120--180 keV.

It is worth noting that the high--energy cutoff is one of the key parameters in reproducing the peculiar ``bump--like''
 shape of the CXB. The ubiquitous presence of an upward curvature (the so--called soft--excess) rising steeply  below 2 keV 
in the spectrum of unabsorbed AGNs is also uncertain (Matt 2000), especially in the case of luminous objects
 (Reeves \& Turner 2000, George et al. 2000).

All the above spectral properties are still to be well constrained for the brightest sources in the local Universe 
and, mostly, there is a evident lack of information about their possible evolution along $z$ and/or luminosity. 

\item {\it The \nh~distribution.} Gilli et al. (1999) and Gilli et al. (2001) assumed in their models a ``fixed''
 distribution of the absorbing 
column densities for Type 2 objects. They used the distribution found by Risaliti et al. (1999) for a sample of 
nearby optically--selected  Seyfert 2 galaxies.
There are three main causes of uncertainty in their assumption: ($i$) the observed distribution is based on a limited luminosity range;
($ii$) it takes into account only ``optical'' Type 2 AGNs i.e. objects 
with narrow emission lines in their optical spectra (but,
as extensively discussed in \p1, a mismatch between optical and X--ray classification of AGN has been
widely observed) and, most important, ($iii$) such distribution has been derived from sources
 in the local Universe (which are clearly not responsible for the bulk of the CXB). Consequently,
 the assumptions made in their model regarding the extension of the \nh~distribution
 to higher redshifts could be erroneous, thus introducing an error in the final output of their estimation.

\item {\it The XLF of Type 2 AGNs.} Recent results from optical follow--ups of {\it Chandra} and \xmm~ deep surveys 
yield a source redshift distribution that, if confirmed, 
would require a substantial revision of the commonly assumed XLFs (Hasinger 2003; Cowie et al. 2003).  

The main uncertainty in all synthesis models of the CXB is indeed the XLF of obscured sources and its cosmological 
evolution, both of
which are completely unknown. So far theoretical models have assumed the same XLF for Type 1 and Type 2 AGNs. 
However, on the basis of the above results (i.e. the bulk of the CXB originates from narrow line AGNs at $z$\simlt~1), 
it has been suggested that the evolution properties of Type 2 sources is likely to be different from that of Type 1s.

Moreover, the commonly--used XLFs of Type 1 AGNs have been mainly derived from soft X--ray surveys and, in addition, 
their evolution with $z$ is still debated (Comastri 2000 for a complete review).

An attempt to adapt the theory to the new observational evidences, has been recently done by
Franceschini, Braito \& Fadda (2002, hereafter FBF02). These authors have proposed a model for the CXB 
where the obscured AGNs closely follow the evolution of strongly evolving infrared starburst galaxies (as recently found by {\it ISO} surveys), 
which evolve steeply up to $z \sim$ 0.8 similarly to the X-ray sources detected in the deep surveys (Cowie et al. 2003). 
This approach is based on the fact that $\sim$ 65\% of the sources (mostly AGNs) detected in the Lockman Hole survey 
in the 5--10 keV band have an IR counterpart at 15 $\mu$m. 
FBF02 therefore suggested two different evolutionary patterns for Type 1 and Type 2 AGNs: the former evolve as given by
optical 
and soft X--ray surveys, whereas the latter evolve faster, as found in  mid--IR surveys.

However, this  scenario seems to be more complicated than described by FBF02 (and subsequently refined by Gandhi \& Fabian 2003):
 in fact, Gilli (2003) has shown that
such a ``starburst--like'' evolution for Type 2 objects largely overestimates the ratio of obscured--to--unobscured objects at $z$ \simlt~1.

\item {\it Space density evolution of different AGN types.} Another hypothesis is
 that classes of AGN with different luminosity may evolve differently, contrary to what
generally assumed in theoretical models, where the evolution is modeled only taking into 
account unabsorbed QSOs. 

In particular, Cowie et al. (2003) and Barger et al. (2003) have recently suggested on 
the basis of \chandra~data that the space density evolution for Seyfert--like objects could be nearly constant
 (or, at most, slightly declining above $z >$ 0.7) up to $z$ $\sim$ 2.5, 
while that of high--luminosity QSO--like AGNs shall increase rapidly 
from $z \sim$ 0 up to $z \sim$ 3 in agreement with the evolution of 
optically--selected samples. 
\end{enumerate}
The need for more accurate XLFs both for obscured and unobscured sources clearly emerges from this
discussion. This information is indeed a crucial input for all models aimed at achieving
a correct and complete interpretation of the CXB phenomenon.

To date, it appears difficult to predict how all this new information obtained by \chandra~and \xmm~will affect synthesis models of the CXB.
It is beyond the goal of this paper to estimate what type of changes these new observational constraints 
(including our new measurement of the fraction of absorbed versus unabsorbed objects) will put on these models, 
but it will be clearly an important development for future theoretical works.
\section{Conclusions}

This work provides the first step in the detailed study of the X--ray spectral properties 
of hard X--ray selected sources detected at faint fluxes 
(i.e. \fhx~\simlt~10$^{-13}$ \cgs) near the knee of the Log$N$--Log$S$
distribution. These are the sources that most contribute to the CXB.

Previously published works on this topic were based  mainly on hardness ratio and/or
stacked spectral analyses. Complementary to ultradeep pencil--beam surveys,
our shallower survey  addresses for the first time the analysis of each individual spectrum. 

Results have been reported for a sample  of 90 hard X--ray selected sources detected
serendipitously in twelve {\it EPIC} fields. This is the largest ever made 
sample of this type. 
A detailed spectral analysis has been performed in order to measure source-by-source the 0.3--10 keV continuum shape, the amount of cold (and, possibly, ionized) absorbing matter and the strength of other spectral features such as soft excess and warm absorber components.\\

The most important results can be  briefly outlined as follows:

\begin{enumerate}

\item Fluxes in the 2--10 (0.5--2) keV band span from 1 (0.04) to 80 (70) $\times$ 10$^{-14}$ \cgs. About
40\% of the X--ray sources are optically classified from the literature. Most of them are broad line AGNs 
with redshift in the range 0.1 \simlt~$z$ \simlt~2. The high luminosities found (10$^{42}$ $\leq L_{2-10} \leq$
 10$^{45}$ \ergs) match well with these identifications except for 
two ``optically dull'' galaxies (see also \p1).

\item Using a  simple power law model with Galactic absorption we obtain $\langle\Gamma\rangle$ = 
1.59$\pm$0.02. Considering only sources in the BRIGHT and FAINT subsamples we find
$\langle\Gamma\rangle$ = 1.53$\pm$0.03 and $\langle\Gamma\rangle$ = 1.56$\pm$0.05, respectively. Both these values are in fairly good agreement with stacked spectral analysis obtained from {\it ASCA} 
and \chandra~hard X--ray surveys.

\item 65\% of the sources are well fitted with the SPL model; their average spectrum provides a photon index
$\Gamma \sim$ 1.7$\div$2.0, i.e. typical of unabsorbed  AGNs.

\item 30\% of the sources require a column density larger than the Galactic value with \nh~ranging
from $\approx$ 10$^{21}$ to $\approx$ 10$^{23}$ \cm2~(Sect.~4). 
In particular, two narrow line AGNs turn out to be Type 2 QSOs 
since they are characterized by high luminosities ($L_{2-10} >$ 10$^{44}$ \ergs) and high column densities (\nh~$>$ 10$^{22}$ \cm2).

\item The mean slope of hard X--ray selected QSOs in our sample remains nearly constant ($\langle\Gamma\rangle \approx$ 1.8--1.9)
between $z \sim$ 0 and $\sim$ 2 (Sect.~\ref{HXSQs}). By combining this result with other 
recent works on high--$z$ QSOs, we strengthen the suggestion 
that QSOs do not exhibit any spectral evolution and, hence,  the type of accretion in these objects should be similar
up to  $z \approx$ 5. 

\item  While from the analysis of the sources
detected in the 1Ms HDFN survey at faint fluxes (\fhx~$<$~10$^{-14}$ \cgs) the  observed fraction of absorbed sources (\nh~\simgt~10$^{22}$ \cm2) is consistent with the theoretical predictions, at the brighter fluxes (\fhx~\simgt~10$^{-14}$ \cgs)
 considered in our survey it appears  to be a factor  $\sim$ 2 lower (with a $\approx$ 4.5$\sigma$ significance) than predicted by synthesis 
models of the CXB. This confirms and extends our previous results obtained in \p1.
\end{enumerate}

\begin{acknowledgements}
This paper is based on observations obtained with \xmm, an ESA science mission with instruments and contributions directly funded by ESA Member States and the USA (NASA).
We would like to thank  Fabrizio Fiore and all the {\it Hellas2XMM} Team for kindly providing optical identifications before publication. 
We also thank Andrea Comastri for providing us his CXB synthesis model in electronic form, and the referee, Roberto 
Gilli, for his careful reading of the paper and comments.
E.P. is greatful to Alessandro Baldi, Christian Vignali and Andrea De Luca for helpful discussions.
This work is partially supported by the Italian Space Agency (ASI). E.P. acknowledges financial support from MIUR for the
Program of Promotion for Young Scientists P.G.R.99.
\end{acknowledgements}

\end{document}